\documentclass[%
preprint,
longbibliography,
amsmath,amssymb,
aps,
pra,
floatfix,
]{revtex4-2}

\usepackage{graphicx}
\usepackage{dcolumn}
\usepackage{bm}
\usepackage{color}

\usepackage[normalem]{ulem}
\usepackage{xcolor}

\begin{document}

\preprint{}

\title{The Footprint of Laminar Separation on a Wall-Bounded Wing Section at Transitional Reynolds Numbers}

\author{Charles J. Klewicki}
\email{charles.klewicki@usc.edu} 
\affiliation{
Department of Aerospace and Mechanical Engineering , 
University of Southern California, Los Angeles, California 90089-1191
}%
\author{Bjoern F. Klose} 
\affiliation{
 Department of Aerospace Engineering, San Diego State University, San Diego, California 92182
}
\author{Gustaaf B. Jacobs}
\affiliation{
 Department of Aerospace Engineering, San Diego State University, San Diego, California 92182
}%

\author{Geoffrey R. Spedding}
\affiliation{
Department of Aerospace and Mechanical Engineering, 
University of Southern California, Los Angeles, California 90089-1191
}

\date{\today}

\begin{abstract}
When a chordwise Reynolds number (Re) falls below about $10^5$ the performance of wings and aerodynamic sections become sensitive to viscous phenomena, including boundary layer separation and possible reattachment. Here, detailed measurements of the flow inside the boundary layer on the suction surface are shown for an aspect ratio 3 wing with wall boundaries.  The separation lines and recirculation zones are shown on the wing and on the wall junction as Re and angle of incidence, ($\alpha$) are varied.  There is good agreement on the lowest Re case which has also been computed in direct numerical simulation. Though the flow at midspan may sometimes be described as two-dimensional, at $\alpha \leq 6^\circ$ it is unrepresentative of the remainder of the wing, and the influence of the wall is seen in strong spanwise flows aft of the separation line. The geometry of the NACA 65(1)-412 section, used here, promotes a substantial chord length for the development of the recirculating regions behind separation making it apt for their study. However, the phenomena themselves are likely to be found over a wide range of wings with moderate thickness at moderate $\alpha$.
\end{abstract}

\maketitle


\section{Introduction}

\subsection{Complex flows at Transitional Reynolds number}
\begin{figure*}
    \begin{center}\includegraphics[width=1\textwidth,]{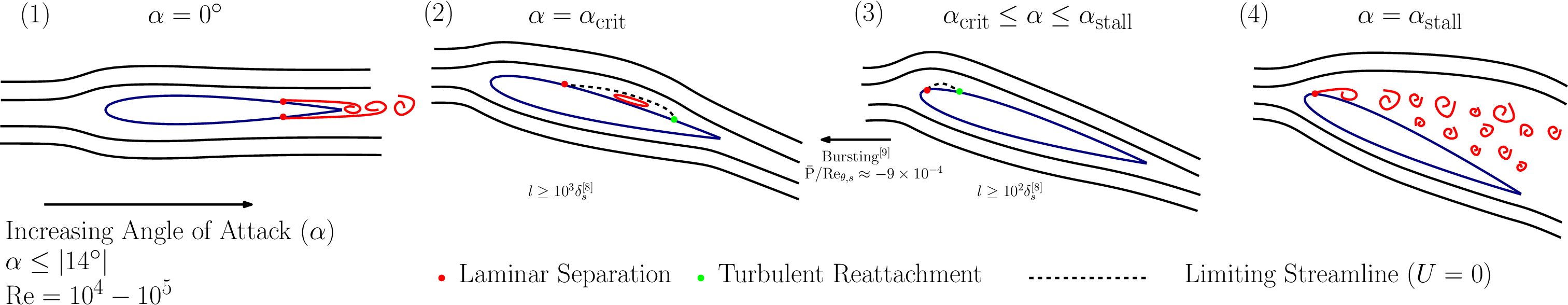}
    \end{center}
    \caption{Some conventional two dimensional flow states as angle of attack is increased for airfoils within Re $ = 10^4-10^5$ regime. (1) trailing edge laminar separation, (2) long laminar separation bubble, (3) short laminar separation bubble, and (4) turbulent separation (stall)}
    \label{fig:Separation States}
\end{figure*}

In the Reynolds numbers range Re $ = 10^4-10^5$ (the chordwise Reynolds number, Re $ = U c/\nu$, where $U$ is the freestream velocity,  $c$ is the chord length and $\nu$ is the kinematic viscosity) the flow on and around an airfoil section or wing is especially complex as the initially laminar boundary layer can separate, transition to turbulence in the detached shear layer, and then in the time-averaged sense, reattach all within one chord length \cite{tank:21, kurelek:21, burgmann:06, burgmann:08a, burgmann:08b, winslow:20}. Separation at these Reynolds numbers is caused by the interaction of the relatively non-robust laminar boundary layer with an adverse pressure gradient created by the curvature of streamlines over the lifting surface. For a given geometry and Re, increasing the angle of attack, $\alpha$, increases the magnitude of the adverse pressure gradient and increases the percentage of chordwise length/area it occupies on the suction surface. The laminar separation point moves towards the leading edge and depending on flow conditions and separation point location, the flow may transition to turbulence and reattach. Laminar separation and time-averaged reattachment is known as the laminar separation bubble (LSB) \cite{tani:64}.Typically four distinct flow states occur for airfoils at Re $ = 10^4-10^5$ and $0^{\circ}\leq\alpha\leq14^{\circ}$: (1) trailing edge laminar separation (without time-averaged reattachment), (2) a long LSB, (3) a short LSB, and (4) turbulent separation, commonly referred to as stall. Tank et al. \cite{tank:21} showed through time-averaged experimental flow fields and force measurements, at matching Re and $\alpha$, that the transition from a flow without reattachment to an LSB occurs at a critical angle of incidence $(\alpha_\text{crit})$. The classification of long and short LSB was proposed by \cite{owen:55} where long LSBs typically have chordwise length $l\geq 10^3\delta_s$ and short LSBs $l\geq 10^2\delta_s$, where $\delta_s$ is the boundary layer thickness at separation. Gaster \cite{gaster:67} proposed that the LSB is governed by Re$_{\theta,s}$ and $\bar{p} = (\theta_s^2/\nu)(\Delta U/ \Delta x)$ where $\theta_s$ is the momentum thickness of the boundary layer at separation and $\Delta U/ \Delta x$ is the change in flow speed over the length of the bubble. A bursting criterion was proposed at $\bar{P}/\mathrm{Re}_{\theta,s} \approx -9\times10^{-4}$ where the bubble switches between the short and long bubble modes. Figure \ref{fig:Separation States} shows a schematic of these different flow states with annotations of how the previous studies relate.

 The flow at transitional chordwise Reynolds numbers does not immediately transition after separation, owing to the stabilizing effects of viscosity. The result is the creation of a shear layer with an initial two dimensional instability, similar to a Kelvin Helmholtz instability with a lower boundary. The two dimensional instability, which may itself be seeded by Tollmein-Schlichting modes \cite{uranga:11, michelis:23} is easily destabilized into a three dimensional flow and is sensitive to small perturbations. Studies focused on acoustic \cite{yang:14, zaman:91, yarusevych:07, golubev:21}, surface \cite{sinha:01, mueller:82} and velocity \cite{tsuchiya:2013, jaroslawski:2023, kay:20, hosseinverdi:2019} perturbations have shown the possibility of suppressing or promoting transition, and in turn modifying airfoil performance. In complement to these experiments, numerical studies \cite{jones:08, jones:10, uranga:11, marxen:10} have shown the complex two- and three-dimensional flow fields resulting from laminar separation and applied linear stability theory to identify instabilities responsible for transition. Linear instabilities present one path and transient energy growth \cite{Trefethen:93} another for the flow to transition into the turbulent regime which, regardless of the exact mechanism, promotes reattachment of the turbulent boundary layer. The dynamics of laminar separation and turbulent reattachment are highly sensitive to $\alpha$, Re, and the flow environment and their locations strongly affect the performance of airfoils \cite{tank:21}. 

\subsection{Influence of Junction}
Many studies of airfoils at Re $ = 10^4-10^5$ invoke two dimensional simplifications, with either periodic boundary conditions in computation or with wall-bounded, spanwise uniform wings in wind tunnel studies. The addition of end walls introduces junction flow phenomena with horseshoe \cite{simpson:01} vortices forming near the leading edge of the airfoil and wrapping around the airfoil creating the possibility of corner separation \cite{dawkins:22,li:20} near the trailing edge and on the wall. Experimental studies \cite{tank:21, pelletier:01, toppings:22} have found that end wall effects for wings with low aspect ratio ($AR = b/c \leq 3$ for a wing of span $b$ and chord $c$) extend considerably into the flow domain. Since wings and turbine blades must be attached to fuselages and rotors in practice, it becomes important to understand the influence of three-dimensional effects and of wall/boundary layer interactions.

\subsection{Thesis}
The purpose of this paper is to experimentally examine the flow topology in the transitional Reynolds number regime of a wall bounded NACA 65(1)412 airfoil. In particular the flow topology of the shearing surfaces will be examined for a wing with AR $=3$. Over a range of chordwise Reynolds numbers $10^4 \leq \textrm{Re} \leq 10^5$ the initially laminar boundary layer always separates and the complex flow structures that emerge after the separation line then appear in a time-averaged sense as an array of laminar separation bubbles of various kinds, depending sensitively on $\alpha$ \cite{klose:25, tank:21}. Sweeps of Re and $\alpha$ will be conducted to study the separation evolution. Comparisons with direct numerical simulations of the flow will be made, with the objective of identifying whether agreement is acceptable for both experiment and simulation to inform control strategies.

\section{Materials and Methods}
\subsection{Water Channel Experimentation}
The USC Blue Water Channel (BWC) is a free surface recirculating water channel with a rectangular test section of dimensions 610 x 890 x 7620 mm$^3$ and turbulence intensity $T \leq 1\%$ over frequencies in the range 0.6 -- 30 Hz  (St $= fc/U = 0.3 - 15$ for $U = 0.4$ m/s) \cite{efstathiou:18},  where $T = (u')_{\text{RMS}}/U$. $T = 1\%$ is close to a threshold for onset of non-modal instabilities reported in \cite{hosseinverdi:2019}, though similarity with companion wind tunnel experiments with more than an order of magnitude decrease in $T$ indicate only a minor influence. The water channel was filled to a height of 500 mm, the highest practical level given physical constraints.
\subsection{Airfoil Model}

\begin{figure}
    \centering
    \includegraphics[width=1\columnwidth,]{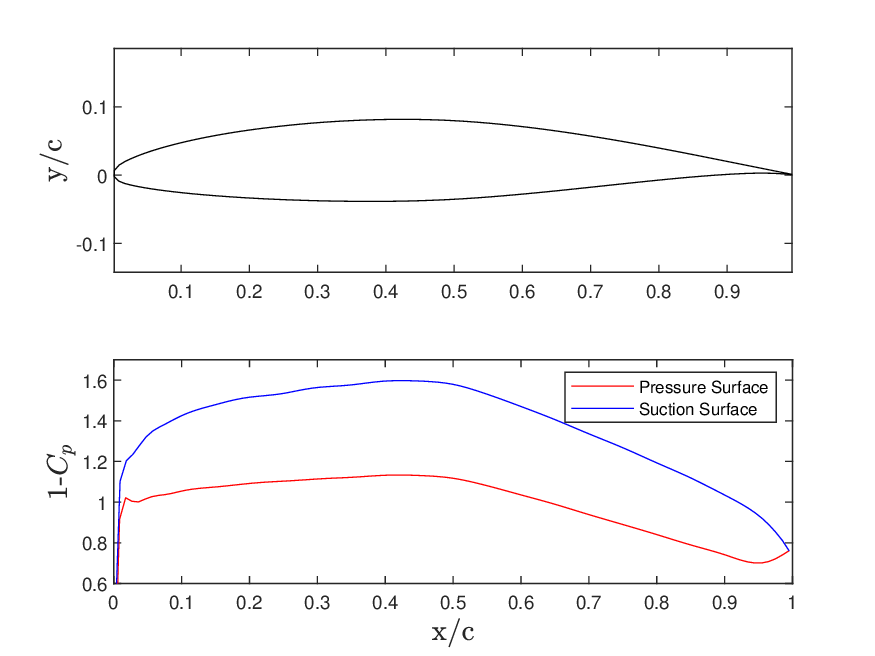}
    \caption{Contour and inviscid pressure profile of NACA 65(1)412 airfoil at $\alpha = 0^\circ$ from in-house panel code.}
    \label{fig:NACA65(1)412}
\end{figure}

A NACA 65(1)412 airfoil was selected as the airfoil model for experimentation and its contour is shown in figure \ref{fig:NACA65(1)412}. The NACA 65(1)412 is used in turbine blades, albeit at significantly higher Re, and was selected here due to its pressure minimum near midchord at $\alpha = 0$, which is well suited to transitional Reynolds number studies, and for the existing companion numerical \cite{klose:25,Klose:21} and experimental \cite{tank:21} studies.

\subsection{Experimental Configuration}

\begin{figure}
    \begin{center}
    \includegraphics[width=0.75\columnwidth,]{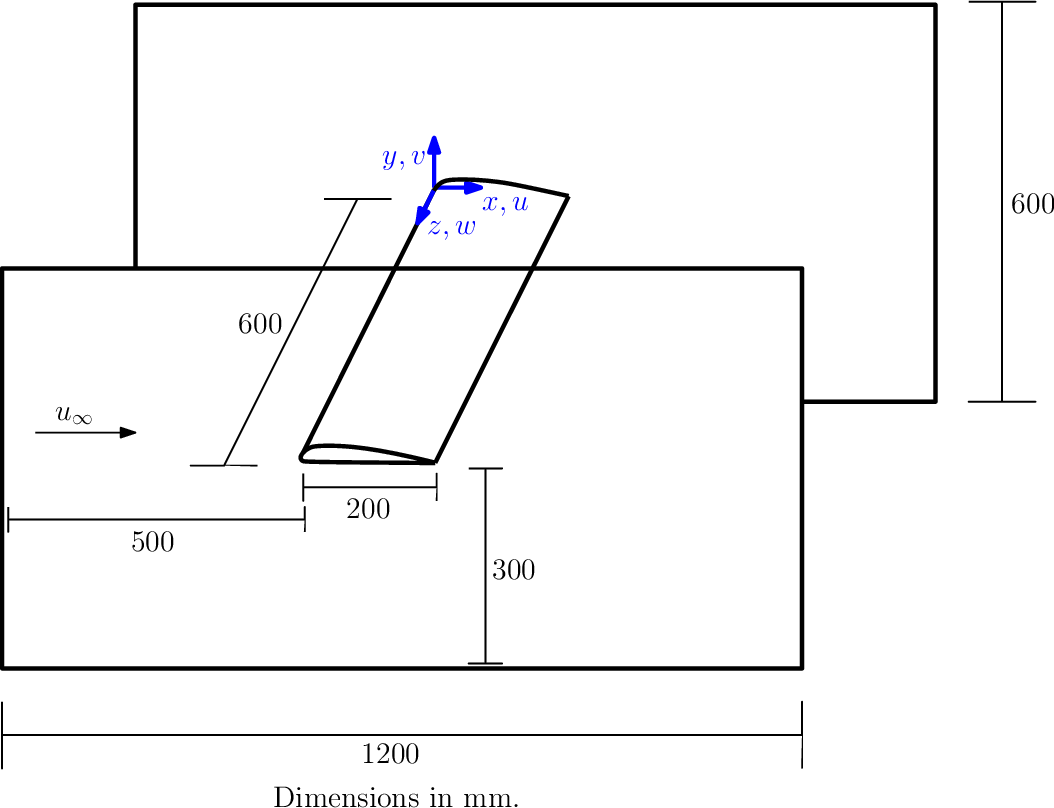}
    \end{center}
    \caption{Schematic of experimental setup in Blue Water Channel.}
    \label{fig:coordinate system}
\end{figure}

The model was 3-D printed using polylactic acid (PLA) material with a chord length of $c = 200$ mm and a span of $b = 600$ mm for an aspect ratio of AR$ = 3$. The airfoil was wet sanded down to 800 grit sand paper and painted black to minimize surface imperfections and reflections. Two transparent acrylic end walls were placed on each side of the airfoil with dimensions of 1200 x 600 x 4.8 mm$^3$. The leading and trailing edges of the end walls were tapered to a point with an angle of 60 degrees. Figure \ref{fig:coordinate system} shows a schematic of the experimental setup. The distance from the leading edge of the end wall to the leading edge of the airfoil at $x = 0$ was 500 mm. 

The airfoil and end walls were fastened together with a corrosion resistant threaded rod and lock washers. $\alpha$ was set using a pin on the side of the airfoil and precision holes drilled into the acrylic which yielded an uncertainty of $\alpha\pm0.5^{\circ}$. The assembly was placed in the water channel 5 m downstream of the inlet with the suction surface of the airfoil facing the bottom of the channel. Using a swivel leveling mount and coupling nut, the threaded rod was put in tension between the walls of the water channel to fix the experimental setup in place. The end walls were intentionally not tripped to imitate the laminar boundary layer developing on the end walls in the complementary numerical study \cite{klose:25}. To reduce the risk of leading edge wall separation swivel leveling mounts were used to slightly incline ($\leq0.49^\circ$ or a Falkner-Skan parameter of $\beta=0.009$) the end walls away from midspan The Reynolds numbers selected for experimentation were Re $= 2, 4, 6, 8 \times 10^4$ which corresponded to free stream speeds of  $U = 0.1, 0.2, 0.3, 0.4$ m/s respectively. Using the laminar flat plate boundary layer thickness estimate, \eqref{deltaFlatPlate} 
\begin{equation}
    \delta = \frac{5x}{\sqrt{Re_x}},
    \label{deltaFlatPlate}
\end{equation}
the boundary layers on end walls were estimated at the airfoil leading edge and are given in table \ref{table:delta end wall}. $\delta_{EW}(x)/c$ is also shown in figures \ref{fig:streamlinesAOA0-2SpanwiseVel}, \ref{fig:streamlinesAOA4-6SpanwiseVel}, and \ref{fig:streamlinesAOA8-10SpanwiseVel} for comparisons with spanwise measurements of the velocity.

A coordinate system with its origin at the end wall and the leading edge of the airfoil was selected with $(x,y,z)$ and $(u,v,w)$ corresponding to the streamwise, chord-normal, and spanwise locations and velocities as shown in figure \ref{fig:coordinate system}.  Velocities for all PIV results were normalized by the freestream: $U$, distances were normalized by the chord length: $c$, and times were normalized by the advection time: $c/U$.

\begin{table}
    \centering
    \begin{tabular}{c|cccc}
         Re $\times 10^{-4}$ & 2 & 4 & 6 & 8 \\
         \hline
         $\delta_{EW}$ (mm) & 11 &  8 & 6.5 & 5.6 \\
         $\delta_{EW}/c$ & 0.055 &  0.040 & 0.033 & 0.028 \\
    \end{tabular}
    \caption{End wall laminar boundary layer thickness estimates at airfoil leading edge.}
    \label{table:delta end wall}
\end{table}

\subsection{2-D Planar Data Acquisition}

\begin{figure}
      \centering
      \includegraphics[width=1\columnwidth]{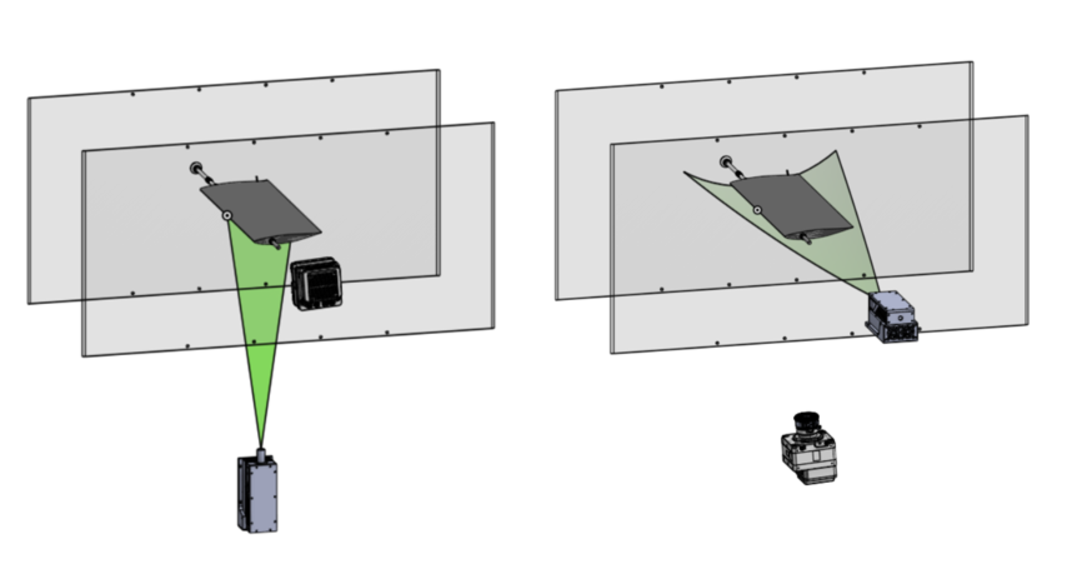}
      \caption{Computer-aided design (CAD) drawing of 2-D laser sheet experimental setup (left). CAD drawing of curved laser sheet experimental setup (right).}
      \label{fig:CAD Water Channel}
\end{figure}

Particle image velocimetry (PIV) experiments were carried out in the water channel using a VEO 410L Phantom Camera and a continuous CNI MGL-N 5 watt 532 nm green laser. The camera resolution was 1280 x 800 pixels with a maximum frame rate, $f_s = 5200 \; \textrm{Hz}$. The channel was seeded with 20 micron polyamide particles. A 100 mm fixed focal length Nikon lens with aperture  f/2.8 was used for a camera field of view (FOV) which captured the entire chord. The PIV results were cropped to extend from the $x/c=0$ (leading edge) to $x/c=1.1$ (just beyond the trailing edge) and from $y/c=0$ (the chord height) to $y/c=0.4$. 1000 images were captured for each data set, equivalent to a minimum of 10 advection times for each Re. Data were captured by fixing $\alpha$ and sweeping through Reynolds number Re = ${2, 4, 6, 8} \times 10^4$ for both the midspan plane and within the end wall boundary layer at $z = 1.5$ mm ($z/\delta_{EW} = 0.14, 0.19, 0.23, 0.27$ for Re = $2,4,6,8\times 10^{4}$ respectively). The experimental setup is shown on the left in figure \ref{fig:CAD Water Channel}.

\subsection{Surface Contoured Data Acquisition}

\begin{figure}
    \begin{center}
    \includegraphics[width=1\columnwidth,]{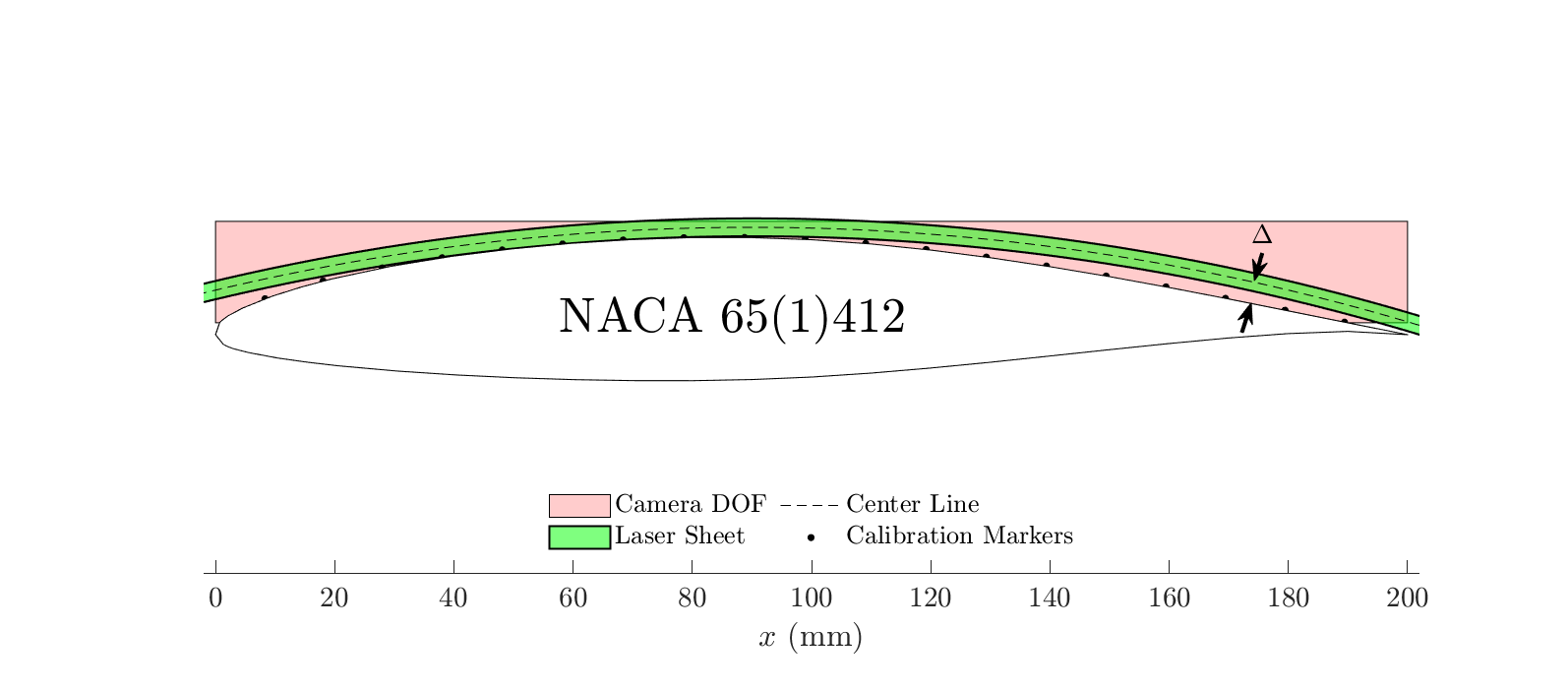}
    \end{center}
    \caption{The geometry of the curved laser sheet, airfoil suction surface and camera depth of field (DOF). $\Delta$ is the distance between the airfoil suction surface and center line of the PIV laser sheet}
    \label{fig:Curved Laser Sheet Schematic}
\end{figure}

In an effort to capture near wall velocities at the suction surface, a laser sheet curved into an arc was generated with a Powell lens and a 1500 mm focal length convex cylindrical lens, by following the techniques of \cite{estevadeordal:10}. The curved laser sheet maintained a thickness of 3 mm while conforming approximately to the airfoil suction surface. A schematic of the curved laser sheet is shown in figure \ref{fig:Curved Laser Sheet Schematic} and table \ref{table:Laser Sheet Distance} gives the estimated distance from the suction surface of the airfoil to the center line of the curved laser sheet. The centerline of the laser sheet is shown in Section \ref{sec:Experimental Results}: figures \ref{fig:midpsanAoA0-2}, \ref{fig:midpsanAoA4-6}, and \ref{fig:midpsanAoA8-10} relative to separated regions on the airfoil at different Re and $\alpha$

\begin{table}
    \begin{center}
        \begin{tabular}{ c|ccccccccccc }
         $x/c$ & 0 & 0.1 & 0.2 & 0.3 & 0.4 & 0.5 & 0.6 & 0.7 & 0.8 & 0.9 & 1 \\
         $x$ (mm) & 0 & 20 & 40 & 60 & 80 & 100 & 120 & 140 & 160 & 180 & 200 \\
         \hline
         $\Delta$ (mm) & 7.48 & 2.2 & 1.58 & 1.50 & 1.60 & 1.94 & 2.66 & 3.36 & 3.70 & 3.40 & 2.16 \\
        \end{tabular}
    \end{center}
    \caption{Distance between the airfoil suction surface and the center of the PIV laser sheet for different chordwise locations.}
    \label{table:Laser Sheet Distance}
\end{table}

\begin{figure}
    \centering
    \includegraphics[width=1\columnwidth,]{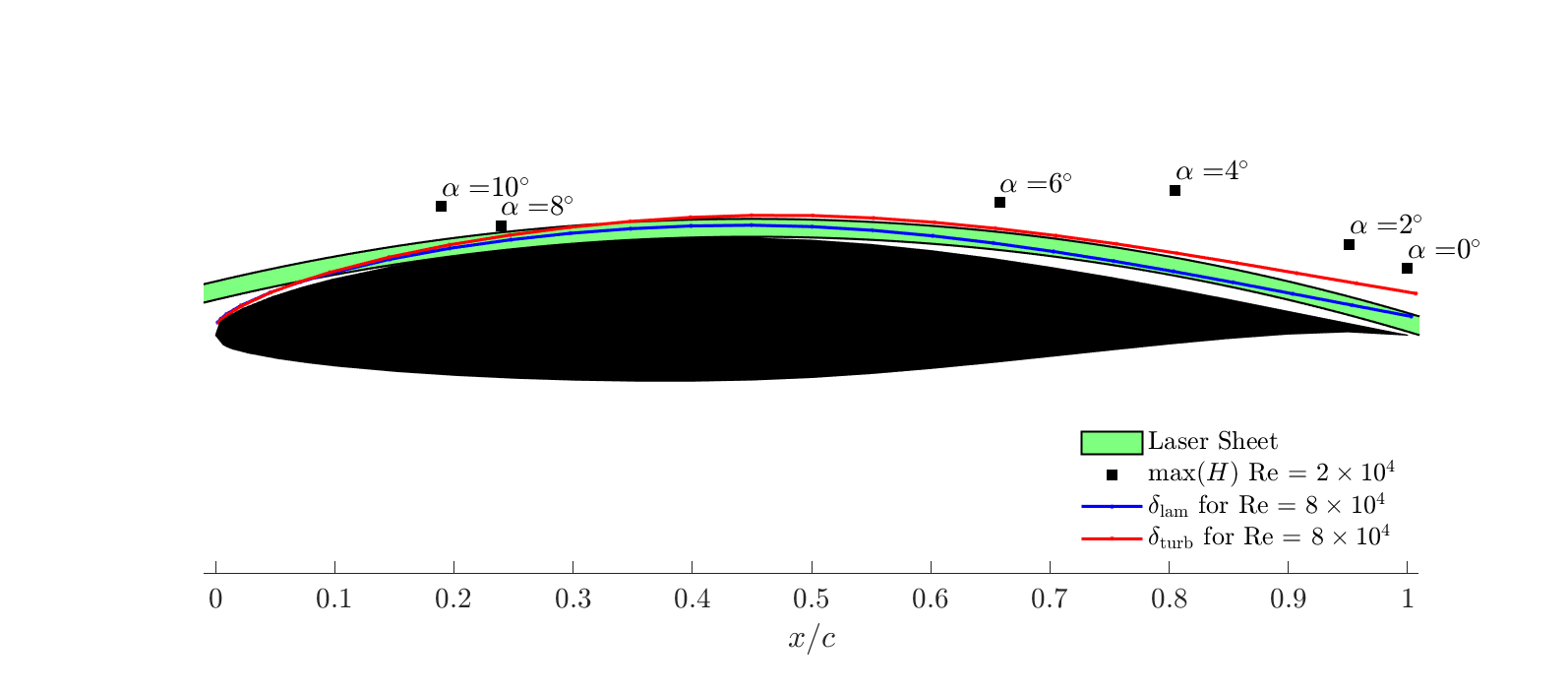}
    \caption{Nondimensional schematic of curved laser sheet with flat plate turbulent and laminar boundary layer thickness estimates for the highest experimental Reynolds number. $(\blacksquare)$ $x/c$ location of maximum shape factor ($H$) plotted at $y=\delta^*$ for Re = $2\times10^4$ across varying $\alpha$. See section \ref{sec:Data Reduction} for details of boundary layer integral parameters.}
    \label{fig:suction surface BL thickness estimates}
\end{figure}

Prior to data acquisition the camera was aligned so the FOV was parallel with the chord. A Bosch GLM50C laser angle measure with an uncertainty of ± 0.2$^\circ$ was used to ensure the camera FOV was parallel with the chord line at each $\alpha$. A 35 mm fixed focal length Nikon lens with an aperture set to f/2 at a distance of 0.83 m was used to capture the entire suction surface of the airfoil. 1000 images were captured for each data set, equivalent to a minimum of 19 advection times for each Re. Experiments were conducted under the same conditions and parameters as the 2-D planar PIV. Figure \ref{fig:CAD Water Channel} shows a CAD drawing of the experimental setup and figure \ref{fig:suction surface BL thickness estimates} shows flat plate boundary layer thickness estimates on the airfoil suction surface for the highest Reynolds number tested as well as the maximum shape factor, plotted at $y=\delta^*$, for the lowest Re. Although the laser sheet is not in contact with the suction surface and small gaps between the two are present, from figure \ref{fig:suction surface BL thickness estimates} it can be concluded that the curved laser sheet will capture dynamics within the boundary layer for locations downstream of $x/c=0.1$ for even the highest Reynolds number case (thinnest boundary layer case).

\subsection{Particle Image Velocimetry: Calibration, Algorithm Parameters, and Uncertainty Estimates}

PIVlab \cite{Thielicke:21} was used to analyze the particle image sequences. The parameters used during analysis are listed in table \ref{table:PIV}. 

\begin{table*}
    \begin{center}
        \begin{tabular}{ |c|c|c| } 
         \hline
         Preprocessing & 
         Correlation & 
         Postprocessing \\ 
         \hline
         CLAHE & FFT Window Deformation & Standard Deviation Validation \\ 
         window size: 64 pixels & 4 passes, 50\% overlap & threshold: 8 \\ 
         \hline
         Weiner2 Filter & Interrogation Areas & Local Median Validation \\
         window size: 2 Pixels & 64, 32, 32, 32 (Planar) & number of neighbors: 4 \\
         &64, 32, 16, 16 (Curved)&\\
         \hline
         Contrast Stretching & Sub pixel Interpolation & Missing Data Interpolation \\ 
         min: 0, max 0.2 & 3 point Gaussian &  Method 4 (inpaint\_nans())\\
         \hline
         Mean Background  & Repeated correlation & -\\ 
         Subtraction & last pass quality slope 0.025 & - \\
         1000 images &  & - \\
         \hline
        \end{tabular}
    \end{center}
    \caption{Parameters used for PIV analysis.}
    \label{table:PIV}
\end{table*}

The pixel displacements were converted to velocities in m/s through
\begin{equation}
    u = \frac{\hat{u}}{\Delta t}F,
    \label{calibration equation}
\end{equation}
where $F=\Delta i$/mm is the conversion factor from pixels to physical units, and $\Delta t = 1/f_s$.

\begin{figure}
    \begin{center}
    \includegraphics[width=0.75\columnwidth,]{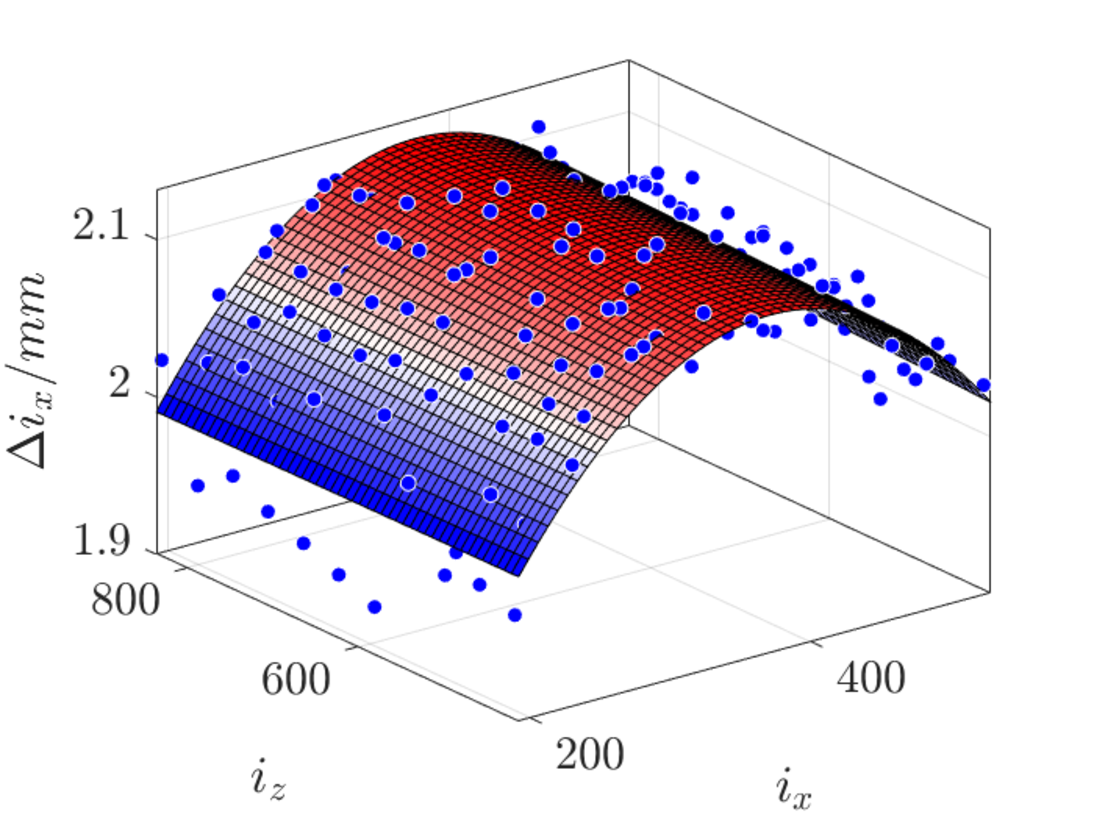}
    \end{center}
    \caption{Chordwise quadratic and spanwise linear polynomial surface fit of streamwise conversion factor $\Delta i_x$/mm as a function streamwise pixel location $i_x$ and spanwise pixel location $i_z$.}
    \label{fig:pixel to mm polynomial fit}
\end{figure}

For the 2-D planar data acquisition the conversion factor was computed by taking an image of a type 11 Lavision calibration plate, at the laser sheet center line location, locating the calibration markers within the image using the MATLAB imfindcircles function, and computing the average pixels per mm from the known physical distance between the markers. After calibration the 2-D planar PIV vector field estimates were rotated to match the coordinate system in figure \ref{fig:coordinate system}. 

Careful consideration was given to the calibration of the curved laser sheet PIV. To convert pixel location/displacements to physical units, a regularly spaced grid of calibration markers was adhered to the suction surface of the airfoil. The calibration grid spanned the middle third of the airfoil and traversed from the leading edge to the trailing edge in the streamwise direction. The calibration marker locations were identified using the same process as the 2-D setup and the pixel distances were computed between markers. Due to the curvature of the laser sheet a single conversion metric was not suitable for calibration. Using the known physical displacement between markers, a conversion factor map as a function of pixel location was created. Chordwise quadratic and spanwise linear polynomial surface fits were computed on the conversion factor map to remove noise and transform the PIV vector field displacements to physical units. The polynomial fits compensated for curvature and perspective effects and are analogous to the magnification factor in \cite{estevadeordal:10}. Figure \ref{fig:pixel to mm polynomial fit} shows an example of a polynomial surface fit produced for calibration. The resulting streamwise vector fields from the PIV analysis represented two-dimensional velocities in the $(x,y)$ plane traveling along the center line of the laser sheet. The streamwise vector fields after conversion were decomposed into the $u$ and $v$ velocities by computing the laser sheet angle relative to the $x$ axis at each vector location and using trigonometric functions to rotate the velocities. 

The PIV analysis yielded a spatial resolution of $x/c \leq 0.02$ for the 2-D planar PIV and curved laser sheet PIV. The different experiments were nearly identical in spatial resolution, since the same particle size and a smaller final interrogation area was used for the larger FOV curved laser sheet data. The temporal resolution of the experiments was resolved to $U\Delta t/c \leq 0.02$. Uncertainty estimates of the time averaged flow speed were computed using Taylor series expansion

\begin{equation}
    \frac{u_\sigma}{U} = \sqrt{\left(\frac{\hat{u}_{\sigma}}{\hat{u}}\right)^2 + 
    \left(\frac{F_\sigma}{F}\right)^2 + 
    \left(\frac{\Delta t_\sigma}{\Delta t}\right)^2
    }.
    \label{uncertainty equation}
\end{equation}

\begin{figure}
    \begin{center}
    \includegraphics[width=1\columnwidth,]{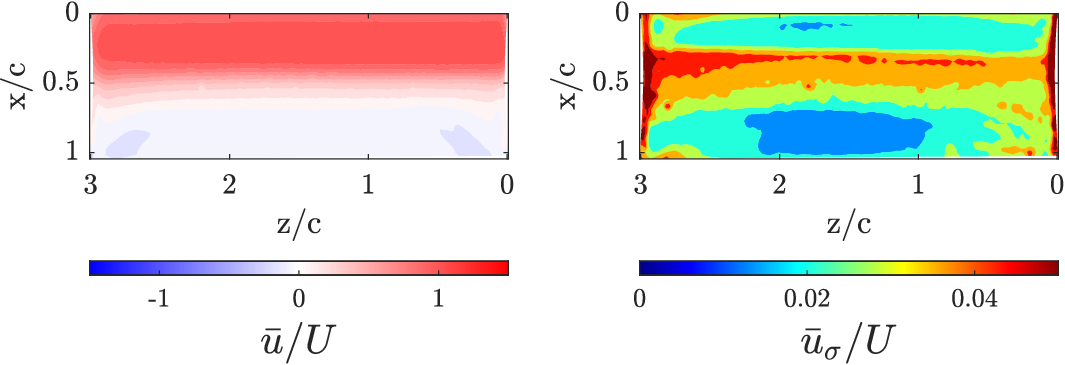}
    \end{center}
    \caption{Representative time averaged streamwise velocity field for the curved laser sheet (left) and its corresponding uncertainty distribution (right).}
    \label{fig:uncertainty field water channel}
\end{figure}

The uncertainty in pixel displacement ($\hat{u}_\sigma$) was computed using the methods of \cite{Sciacchitano:13}, the uncertainty in conversion factor ($F_\sigma$) was taken as the standard deviation about the polynomial fit, and the uncertainty in displacement time ($\Delta t_\sigma$) was assessed to be negligible from the use of a continuous laser.
Figure \ref{fig:uncertainty field water channel} shows a representative time averaged velocity field for the curved laser sheet and its corresponding time averaged uncertainty field ($\overline{u}_\sigma$). The uncertainty field shows that for all locations the uncertainty was below $0.1U$ and the majority of the flow had uncertainties between $0.02U$ and $0.04U$. Uncertainties increased in areas where high spatial gradients occurred such as the $x/c=0.4$ line and areas of high reflections (e.g. the end plates at $z/c=0, 3$)  

\subsection{Data Reduction}
\label{sec:Data Reduction}
Data reduction of the PIV results was accomplished by time averaging, and these quantities are denoted by an overbar (e.g. $\bar{u}$). Kinetic energies of the flow velocities were investigated in section \ref{sec:Experimental Results} and were computed by first decomposing each velocity component $(u, v, w)$ into its temporal mean and fluctuating part
\begin{equation}
    u' = u - \bar{u},
    \label{eqn:velocity decomposition}
\end{equation}
and then computing the kinetic energy associated with the fluctuating components
\begin{equation}
    K = \frac{1}{2}\left(\overline{(u')^2} + \overline{(v')^2}+ \overline{(w')^2}\right).
    \label{eqn:Fluctuating Kinetic Energy}
\end{equation}

Boundary layer integral parameters were computed using equations:

\begin{equation}
    \delta^* = \int^\delta\left(1-\frac{u}{u_e}\right)dy',
    \label{eqn:delta star}
\end{equation}

\begin{equation}
    \theta = \int^\delta \frac{u}{u_e}\left(1-\frac{u}{u_e}\right)dy',
    \label{eqn:theta}
\end{equation}

\begin{equation}
    H = \delta^*/\theta.
    \label{eqn:H}
\end{equation}

\begin{figure}
    \begin{center}
    \includegraphics[width=1\columnwidth,]{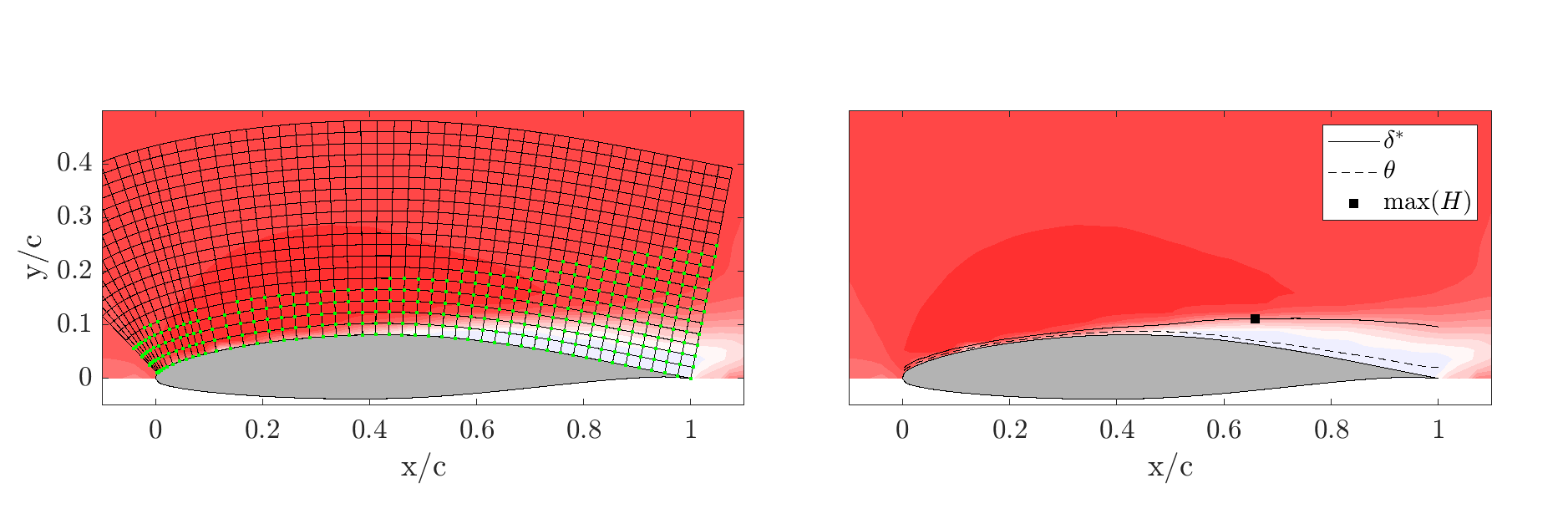}
    \end{center}
    \caption{(Left): Interpolation of curvilinear coordinate system onto streamwise velocity contour. The black grid is the uniform surface-normal height coordinate system. The green points are the grid points extending from the surface ($y'=0$) to the edge of the boundary layer ($y'=\delta$) at each surface-normal location. (Right): Streamwise velocity contour with boundary layer integral parameters.}
    \label{fig:BLIntegralParameters}
\end{figure}

\begin{figure}
    \begin{center}
    \includegraphics[width=1\columnwidth,]{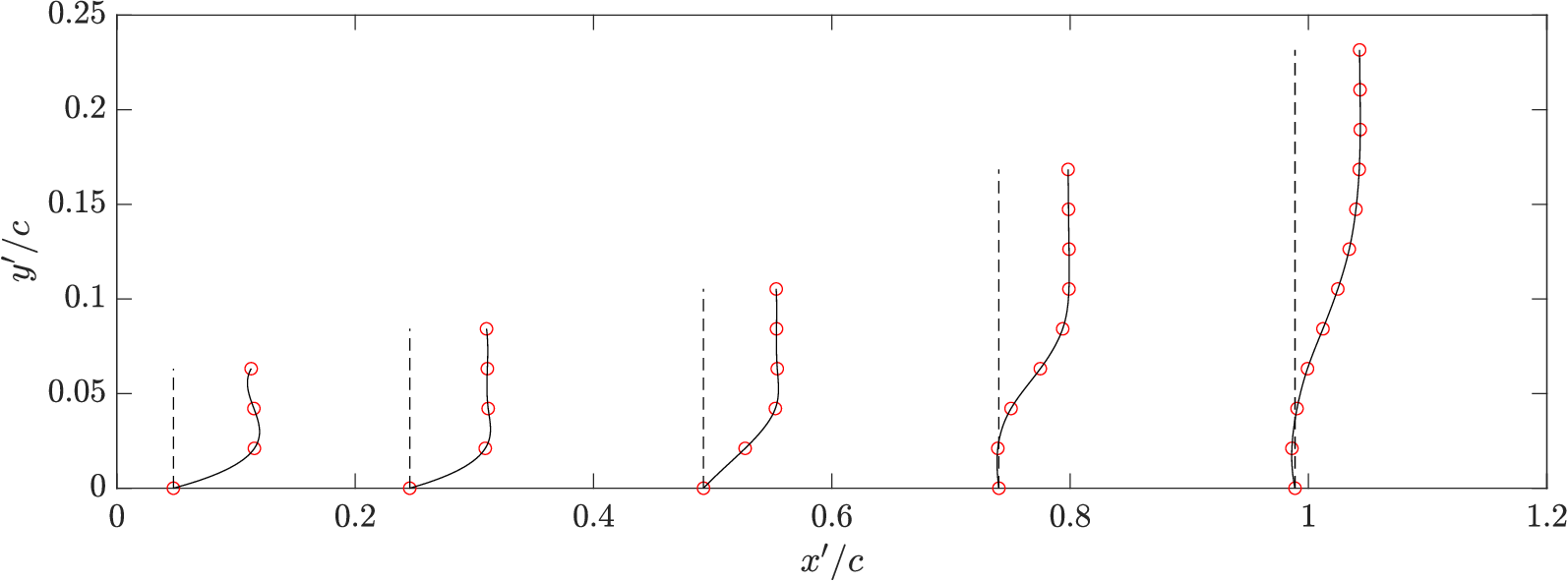}
    \end{center}
    \caption{Velocity profiles at different streamwise locations. ($--$): u=0, ($-$): cubic spline fit, (circles): PIV data points.}
    \label{fig:BLIntegralVelocityProfiles}
\end{figure}

Prior to computation of the boundary layer integral parameters a transformation from the chord based coordinate system $(x,y)$ to the (suction) surface-normal coordinate system $(x',y')$ was performed as shown in figure \ref{fig:BLIntegralParameters}. The grid of the transformation was kept consistent with the PIV resolution ($\Delta x, \Delta y =0.02c$) and careful consideration was given to the effects of the limited PIV spatial resolution. At each surface location the edge of the boundary layer $\delta=y(u_e=\max u)$ was determined and to mitigate resolution concerns a cubic spline was fit to the data at each chordwise location as shown in figure \ref{fig:BLIntegralVelocityProfiles}. A minimum of 4 PIV grid points was used for the cubic splines. Using the spline fits $\delta^*$, $\theta$, and $H$ were computed for Re=$2\times10^4$ and $\alpha=0^\circ-10^\circ$

It should be noted that these boundary layer parameters are inherently non-unique results due their integral nature (i.e. the same value could be obtained from a different flow at a different $x'$ location with a different pressure distribution), but are useful for quantitative comparisons. In particular, shape factor $H$ has been shown to be a practical measure of the separation location and the onset of transition for flows with LSBs \cite{kurelek:18,burgmann:08b,uranga:11,klose:25}. Only the parameters for Re $=2\times 10^4$ were computed due to the spatial resolution limitations of the flows at higher Re. Table \ref{table: H} shows the shape factor of PIV measurements as a function of $x/c$ for different $\alpha$ with the maximum value for each row in bold font. Table \ref{table: H} is a coarse grid and applying a finer grid as shown in figure \ref{fig:suction surface BL thickness estimates} gives better estimates of the $x/c$ location of $\max(H)$.

\begin{table}
    \begin{center}
        \begin{tabular}{ c|cccccccccc } 
         $x/c$ & 0.1 & 0.2 & 0.3 & 0.4 & 0.5 & 0.6 & 0.7 & 0.8 & 0.9 & 1 \\
         \hline
         $\alpha=0^\circ $ & 2.38 & 2.33 & 2.45 & 2.23 & 2.25 & 2.27 & 2.49 & 4.08 & 5.44 & \textbf{6.61}  \\
         $\alpha=2^\circ $ & 2.39 & 2.34 & 2.39 & 2.25 & 2.24 & 2.33 & 3.57 & 4.96 & 7.45 & \textbf{8.74} \\
         $\alpha=4^\circ $ & 2.43 & 2.33 & 2.35 & 2.26 & 2.86 & 6.32 & 7.28 & \textbf{13.1} & 11.6 & 9.67  \\
         $\alpha=6^\circ $ & 2.26 & 2.41 & 2.26 & 2.36 & 2.93 & 4.98 & 6.12 & \textbf{6.22} & 5.57 & 4.74  \\
         $\alpha=8^\circ $ & 2.11 & 2.57 & \textbf{2.73} & 2.58 & 2.28 & 2.03 & 2.07 & 2.12 & 2.16 & 2.17 \\ 
         $\alpha=10^\circ $ & 4.81 & \textbf{4.97} & 4.32 & 3.31 & 2.95 & 2.63 & 2.72 & 2.77 & 3.04 & 3.09 \\ 
        \end{tabular}
    \end{center}
    \caption{Shape factor $H(x)$ for Re $=2\times 10^4$ PIV measurements.  Maximum $H$ are shown in bold for each $\alpha$.}
    \label{table: H}
\end{table}

\begin{figure}
    \begin{center}
    \includegraphics[width=1\columnwidth,]{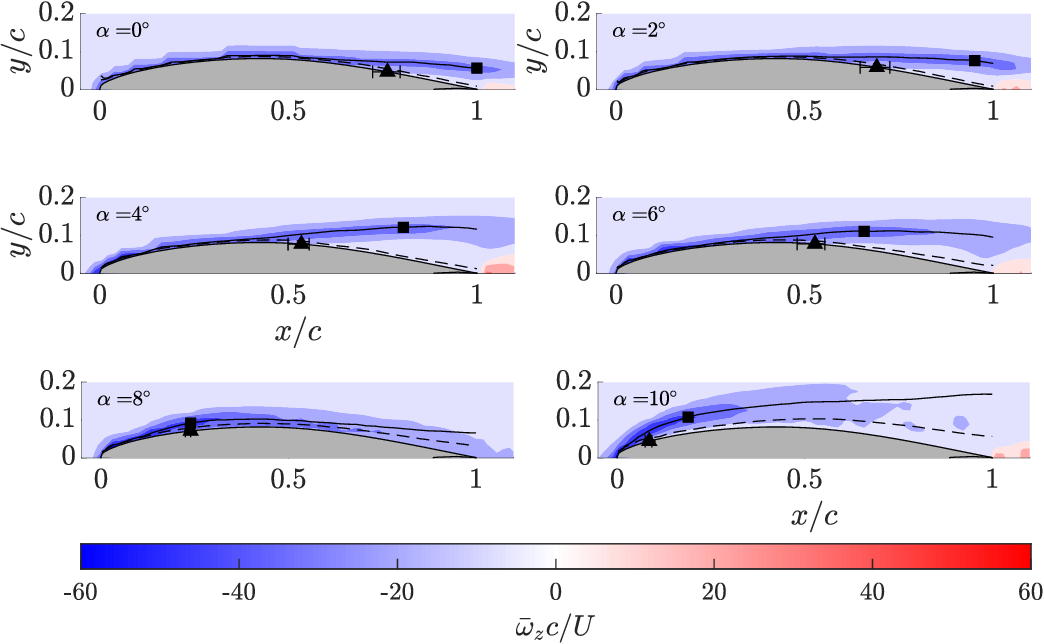}
    \end{center}
    \caption{Midspan $\overline{\omega}_z$ contours with boundary layer integral parameters  $(-)$: $\delta^*$, $(--)$: $\theta$, $(\blacksquare)$: $(\max H)$, and separation location $(\blacktriangle)$: $x_s$ for Re $=2\times 10^4$. $x_s$ uncertainty bars represent location of the shape factor corresponding to the Blasius boundary layer solution and the Falkner-Skan boundary layer solution at separation.}
    \label{fig:SeparationEstimatesRe=20k.eps}
\end{figure}

The shape factor estimates can be used to estimate a separation location ($x_s$). $x_s$ is estimated based on the shape factor of two canonical steady boundary layers: the Blasius boundary layer acting as a lower limit ($H=2.59$) and the Falkner-Skan boundary layer at separation ($\beta=-0.2$) acting as an upper limit ($H=3.95$). These flows are both assumed to be fully developed, while laminar airfoil flows are streamwise developing flows and likely separate somewhere between these values. Here, $x_s$ is naively identified as the location that bisects these bounds $H=3.27$, although Kurelek et al \cite{kurelek:18} showed $H\approx4$ near $x_s$. Figure \ref{fig:SeparationEstimatesRe=20k.eps} shows $x_s$ as a function of $\alpha$ at the midspan plane in relation to the other boundary layer parameters and superimposed on $\overline{\omega_z}$ as computed from equation \eqref{eqn:omega_z}. $\delta^*$ can be observed to be representative of the shear layer location post separation and $x_s$ appears close to the intersection of the high vorticity contour with the suction surface. The $\alpha=8^\circ$ stands out as $x_s (\blacktriangle) \approx\max H (\blacksquare)$. The separation estimate for this $\alpha$ is likely poor due to spatial resolution limitations. These estimates will be used in section \ref{sec:Results} to relate the separated regions to the different PIV planes and separation structures.

Temporal variations of the flow at midspan were analyzed through plots of the z-component of vorticity 
\begin{equation}
    \omega_z = \frac{\partial v}{\partial x} - \frac{\partial u}{\partial y}
    \label{eqn:omega_z}
\end{equation}
and power spectral density (PSD).
The power spectral density ($\hat{P}_v$) of the vertical velocity component $v$, at a location in the shear layer was found by first computing the one-sided Discrete Fourier Transform (DFT) given by

\begin{equation}
    V_k = \sum_{n=0}^{(N-1)/2} 2 v_n e^{-i 2 \pi n k/N},
    \label{eqn:DFT}
\end{equation}
where $v_n$ are velocity values at discrete times, $v(t^n)$, $N$ is the number of samples and $k/N$ is the sinusoidal frequency. The power spectral density is then
\begin{equation}
    \hat{P}_v= |V_k|^2/(F_sN),
    \label{eqn:PSD}
\end{equation}
where $F_s$ is the sampling frequency. Frequencies were nondimensionalized by St$=fc/U$. The $(x,z)$ location in the shear layer of the $v(t)$ trace for each angle of attack was selected based on $\overline{\omega}_z$ contours, and is shown by magenta markers in figure \ref{fig:AoA06PSD} and \ref{fig:Re20000PSD}.

Phase averaging is denoted with a tilde (e.g. $\tilde{u}$),
\begin{equation}
    \tilde{u}(\phi) =  \frac{1}{N_{\phi}}\sum_i^{N_{\phi}} u(\phi_0+Ti)
\end{equation}
where $T$ is the period of phase averaging, $\phi$ is the phase instance, $\phi_0 = \phi(t_0)$, and $N_{\phi}$ is the number of complete periods captured in the data.

\section{Results}
\label{sec:Results}
\subsection{PIV Time Averaged Flow Fields} \label{sec:Experimental Results}
\subsubsection{Laminar separation at low $\alpha$}
\label{sec:Laminar separation at low alpha}

\begin{figure}[hb]
    \centering
    \includegraphics[width=1\columnwidth,]{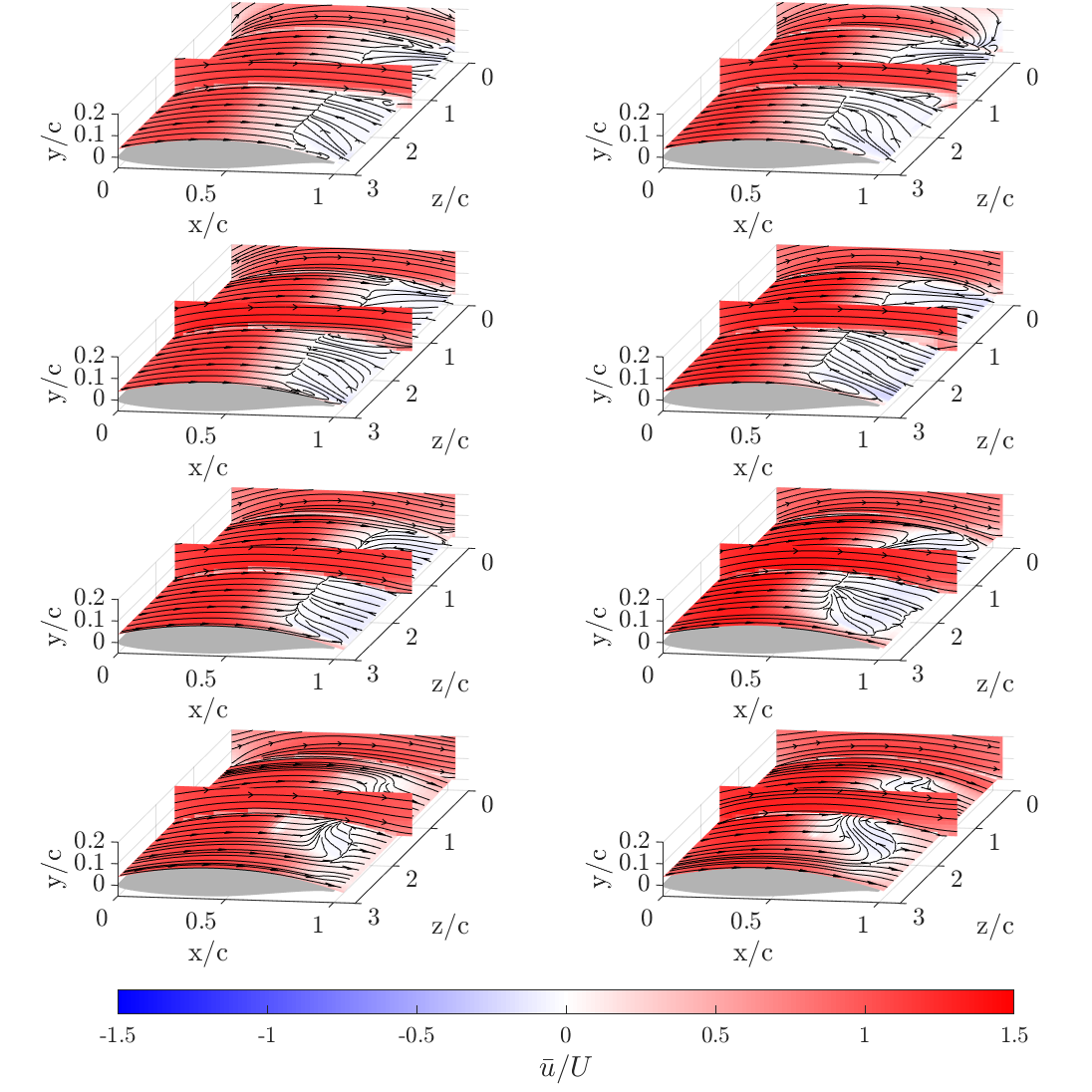}
    \caption{Time averaged streamlines and $\overline{u}$ for $\alpha = 0^{\circ}$ (left) and $ 2^{\circ}$ (right). Re $= 2,4,6,8 \times 10^4$ (top to bottom).}
    \label{fig:streamlinesAoA0-2}
\end{figure}

\begin{figure}
    \centering
    \includegraphics[width=1\columnwidth,]{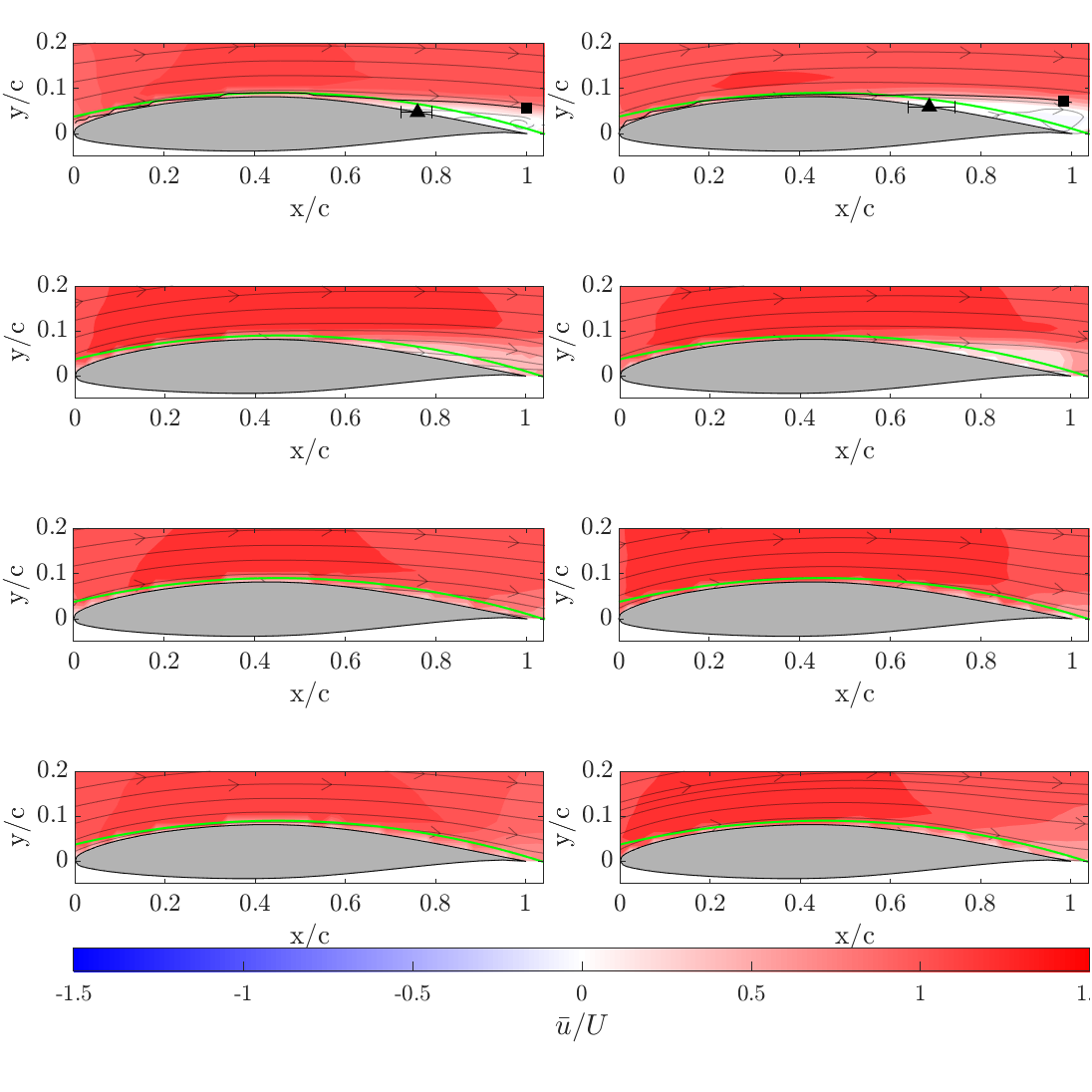}
    \caption{Time averaged streamlines and $\overline{u}(x, y)$ for $\alpha = 0^{\circ}$ (left) and $ 2^{\circ}$ (right) at midspan. Re $= 2,4,6,8 \times 10^4$ (top to bottom). The green line is the centerline location of the curved laser sheet over which the spanwise distributions in figure \ref{fig:streamlinesAOA0-2SpanwiseVel} are estimated. Boundary layer integral parameters $(-)$: $\delta^*$, $(\blacksquare)$: $(\max(H))$, and separation location $(\blacktriangle)$: $x_s$  error bars are based on $H=2.59-3.95$}
    \label{fig:midpsanAoA0-2}
\end{figure}

\begin{figure}
    \centering
    \includegraphics[width=1\columnwidth,]{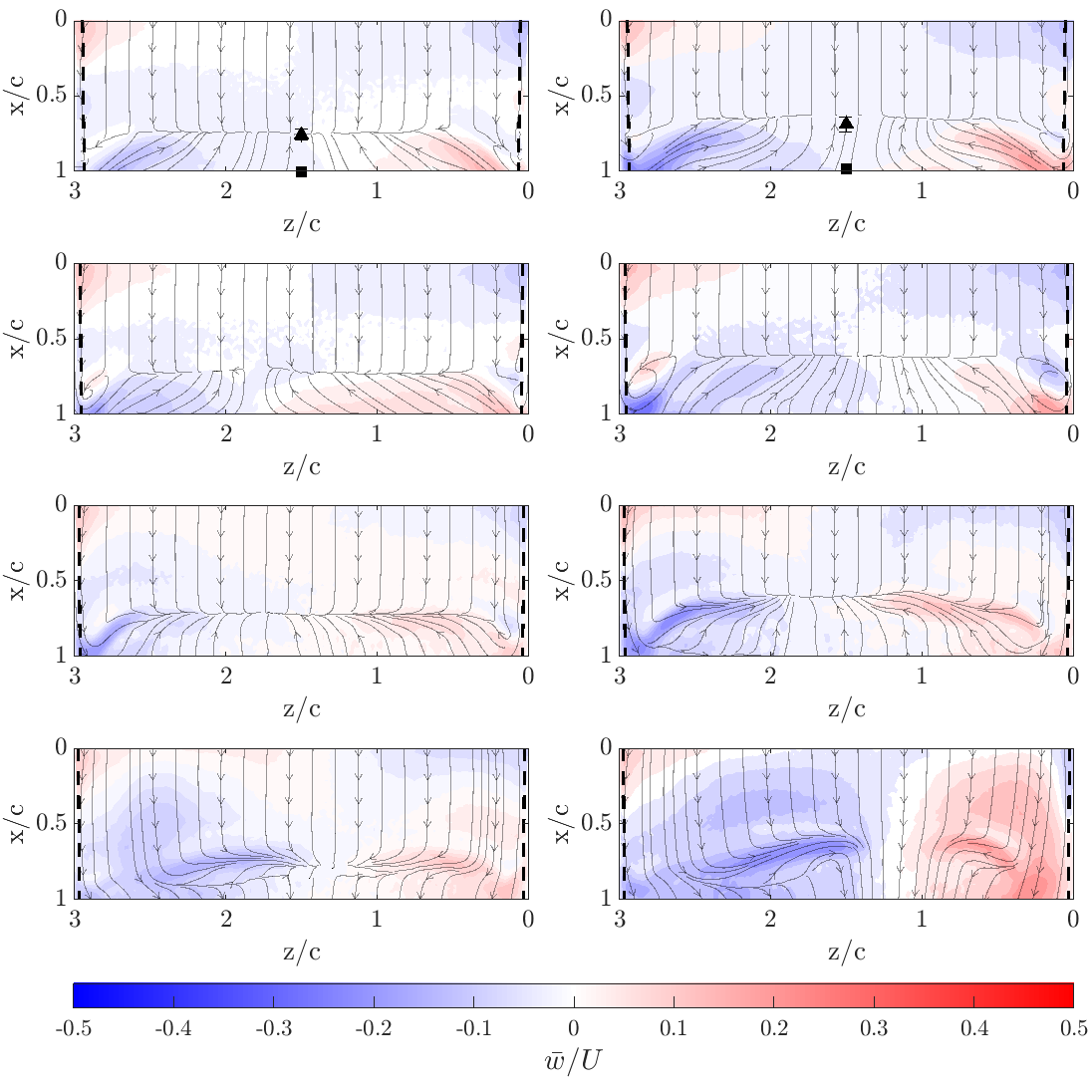}
    \caption{$\overline{w}(x,z)$ for $\alpha = 0^{\circ}$ (left) and $2^{\circ}$ (right). Re $= 2,4,6,8 \times 10^4$ (top to bottom). ($--$): Estimate of laminar boundary layer thickness ($\delta_{EW}(x)$) on the end walls, $(\blacksquare)$: $(\max(H))$,  $(\blacktriangle)$: $x_s$. $x_s$ error bars are based on $H=2.59-3.95$}
    \label{fig:streamlinesAOA0-2SpanwiseVel}
\end{figure}

\begin{figure}
    \centering
    \includegraphics[width=1\columnwidth,]{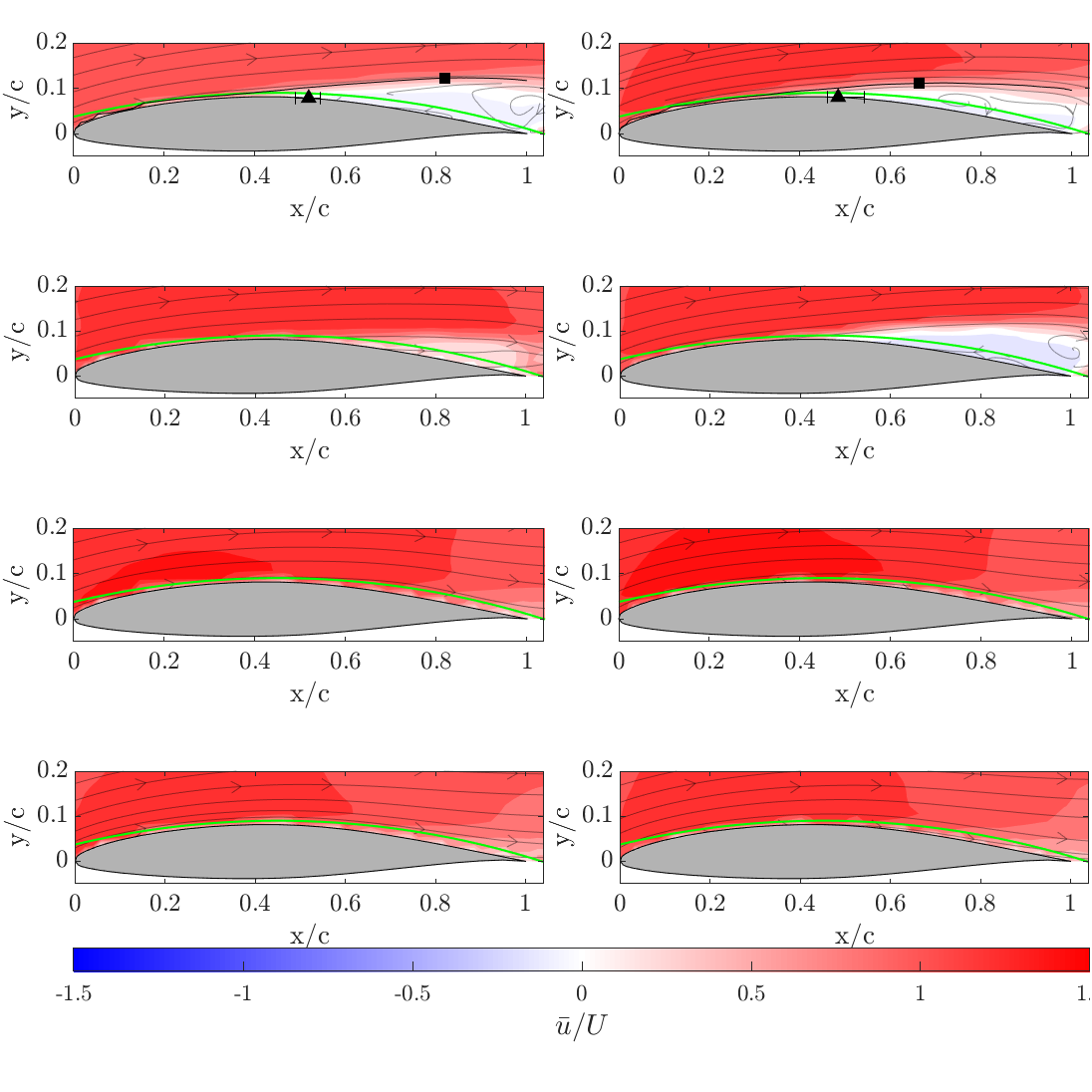}
    \caption{Time averaged streamlines and $\overline{u}(x, y)$ for $\alpha = 4^{\circ}$ (left) and $ 6^{\circ}$ (right) at midspan. Re $= 2,4,6,8 \times 10^4$ (top to bottom). The green line is the centerline location of the curved laser sheet over which the spanwise distributions in figure \ref{fig:streamlinesAOA0-2SpanwiseVel} are estimated. Boundary layer integral parameters $(-)$: $\delta^*$, $(\blacksquare)$: $(\max(H))$, and separation location $(\blacktriangle)$: $x_s$. $x_s$ error bars are based on $H=2.59-3.95$}
    \label{fig:midpsanAoA4-6}
\end{figure}

Figures \ref{fig:streamlinesAoA0-2} - \ref{fig:streamlinesAOA0-2SpanwiseVel} show the time averaged velocity fields for $\alpha = 0^{\circ}$ and $\alpha=2^{\circ}$. In both cases the airfoil experiences laminar separation over the second half of the chord with a separated shear layer and flow reversal beneath. The midspan planes at Re  $= 2 \times 10^4$ demonstrate these phenomena most clearly and also show as $\alpha$ is increased from $0^{\circ}$ to $2^{\circ}$ the $\overline{u} = 0$ line in the $(x,y)$ plane moves towards the leading edge. The midspan planes show a decreasing angle between the $\overline{u} = 0$ line and the suction surface as Re is increased. This angle flattens below the resolution of the experimental measurements at Re $= 6 \times 10^4$, as shown in figure \ref{fig:midpsanAoA0-2}

The suction surface streamlines in figure \ref{fig:streamlinesAOA0-2SpanwiseVel} show that the $\overline{u} = 0$ location moves towards the leading edge in the $(x,z)$ plane as $\alpha$ increases from $0^{\circ}$ to $2^{\circ}$. We shall refer to this line at $\overline{u}=0$, between forward and reversed flow regions, as the separation line denoted $\overline{u}_0$. Technically, the separation point in a two-dimensional flow is diagnosed not through zero-crossings of the streamwise velocity, but of the time averaged skin friction coefficient \cite{haller:04}.  In the absence of gradient information on the wall, the $\overline{u}_0$ criterion is used as an approximation. Re has little effect on the $(x,z)$ location of the $\overline{u}_0$ line at the root of the foil for Re $= 2,4,6 \times 10^4$, but at Re $ = 8 \times 10^4$ the separation region near the root collapses. At the same time the separated region close to midspan also closes, and the equivalent frames in figure \ref{fig:streamlinesAoA0-2} confirm that the flow is fully attached in both locations.  If the reason is the development of a turbulent boundary layer close to the leading edge which then causes a thin separation bubble, then this development is not uniform across the span. At all Re, the surface flow downstream of the separation line has a significant spanwise component, exceeding the expected influence of end wall boundary layer growth (shown by the dashed line in figure \ref{fig:streamlinesAOA0-2SpanwiseVel}), which brings fluid from the wall towards midspan. The flow field is not two-dimensional anywhere on the suction surface, except arguably at the centerline symmetry point (which itself drifts back and forth in $z$).  The spanwise flow brings the influence of the wall to cover most of the span.  At the higher Re, there is no reversed flow towards the trailing edge but the strong spanwise flow is focused at the location where the separation line last occurred.  Note that because the curved light sheet is offset further from the surface (see figure \ref{fig:midpsanAoA0-2}), though still inside the boundary layer, it is possible that forward or backward motion still occurs at the surface, beneath the interrogation slice.

\begin{figure}
    \centering
    \includegraphics[width=1\columnwidth,]{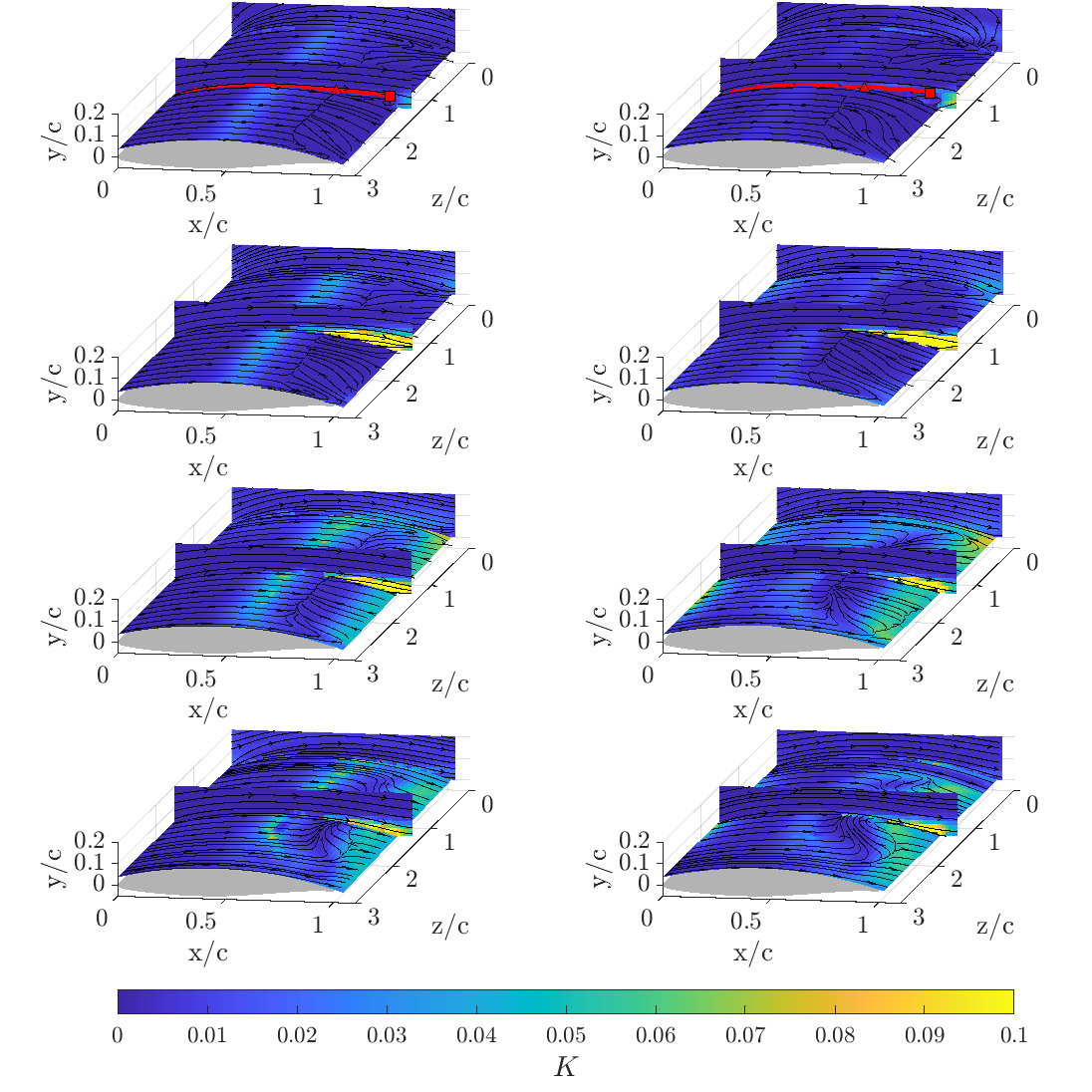}
    \caption{Kinetic energy of fluctuations, $K$, for $\alpha = 0^{\circ}$ (left) and $ 2^{\circ}$ (right). Re $= 2,4,6,8 \times 10^4$ (top to bottom). $(-)$: $\delta^*$, $(\textcolor{red}{\blacksquare})$: $\textcolor{red}{\max(H)}$.}
    \label{fig:streamlinesAOA0-2FlucEnergy}
\end{figure}

Figure \ref{fig:streamlinesAOA0-2FlucEnergy} shows the time averaged kinetic energy of fluctuations mapped onto the principal interrogation planes and the same streamline pattern as figure \ref{fig:streamlinesAOA0-2SpanwiseVel}. At $\alpha = 0^\circ$, the most striking feature is a light band of higher fluctuation amplitudes that spans the whole wing at about midchord. This band becomes significantly distorted only at the higher Re $= 8 \times 10^4$, and it appears always upstream of the separation line. The increase of fluctuations upstream of $\bar{u}_0$ is consistent with the onset of linear instability occurring upstream of separation shown by previous studies \cite{michelis:17, diwan:09, yeh:20}. The light band is surrounded by the dark blue of low $K$, laminar flow all over the remainder of the suction surface, and the only bright spots lie downstream of the trailing edge. The light regions likely come from quasi-periodic fluctuations from the trailing edge, and from the movement of a region upstream of the separation line. The dashed red line and red dot indicate $\delta^*$ and $\max(H)$ at Re $ = 2 \times 10^4$. The flow is not reattached at this Re so $\max(H)$ is a poor estimate of transition. As would be expected, $\delta^*$ closely follows the airfoil surface until reaching the shear layer at which point it diverges from the surface and follows the curve of the shear layer until the trailing edge. At Re $ = 4 \times 10^4$ (second row in the figure), the separated region has lit up in the midspan plane but not on the airfoil suction surface which remains laminar.  With further increases in Re $ = 6, 8 \times 10^4$ (third and fourth row) increased fluctuation levels are observed in the separated flow region.

\subsubsection{Strongly three-dimensional, partly re-attached flows at intermediate $\alpha$}

\begin{figure}
    \centering
        \includegraphics[width=1\columnwidth,]{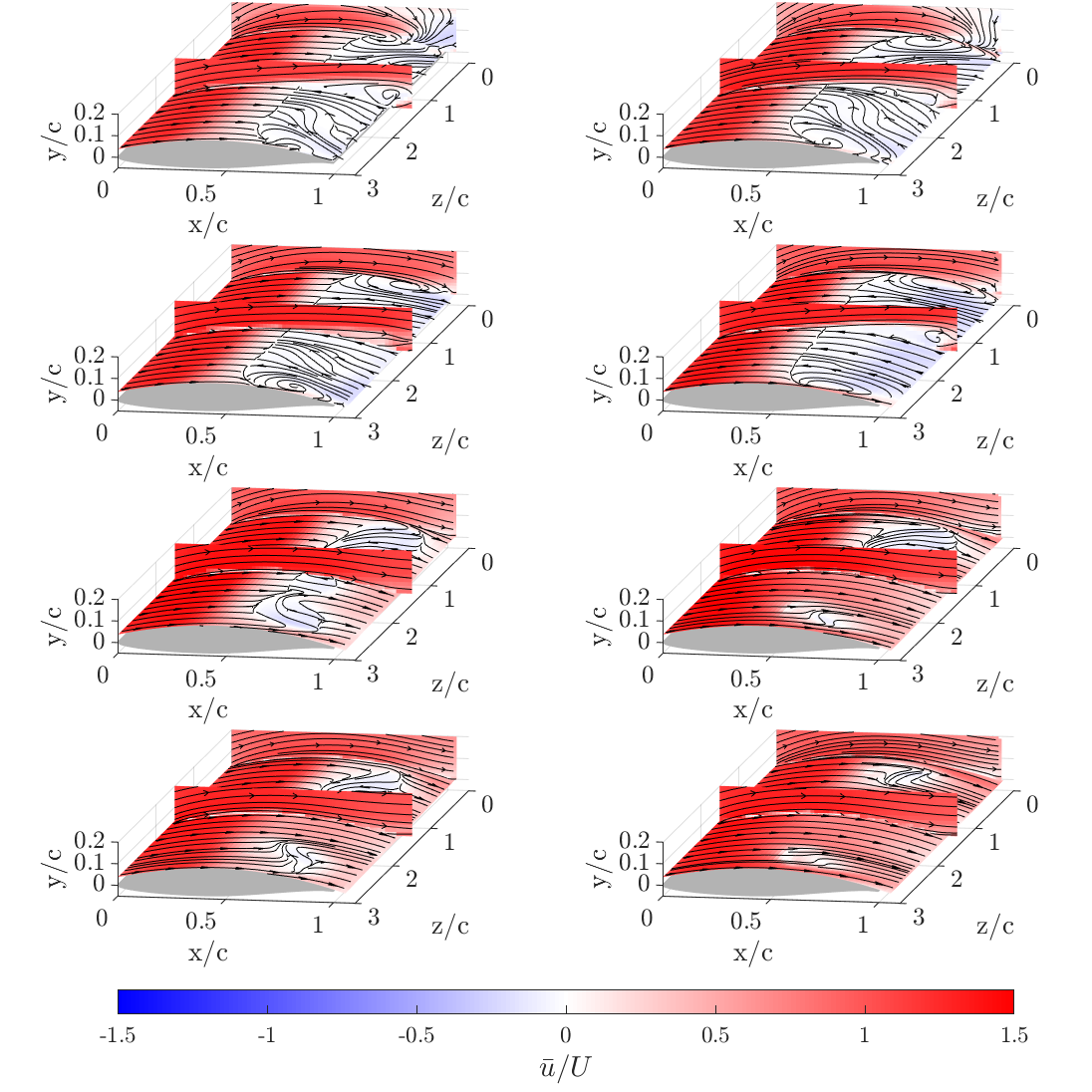}
    \caption{Time averaged streamlines and $\overline{u}$ for $\alpha = 4^{\circ}$ (left) and $ 6^{\circ}$ (right). Re $= 2,4,6,8 \times 10^4$ (top to bottom).}
    \label{fig:streamlinesAOA4-6}
\end{figure}

\begin{figure}
    \centering
    \includegraphics[width=1\columnwidth,]{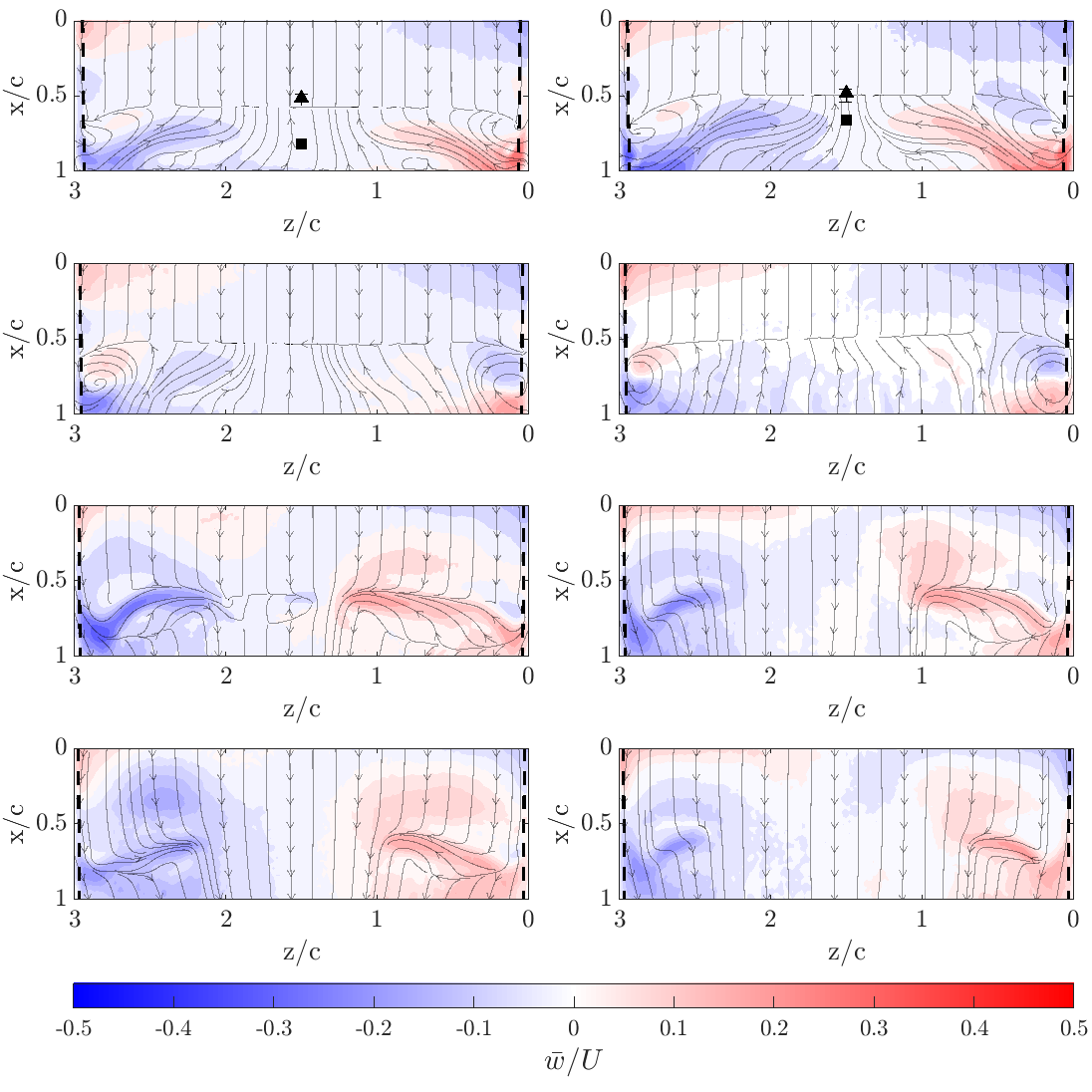}
    \caption{$\overline{w}(x,z)$ for $\alpha = 4^{\circ}$ (left) and $6^{\circ}$ (right). Re $= 2,4,6,8 \times 10^4$ (top to bottom). ($--$): estimate of laminar boundary layer thickness ($\delta_{EW}(x)$) on the end walls, $(\blacksquare)$: $(\max(H))$,  $(\blacktriangle)$: $x_s$. $x_s$ error bars are based on $H=2.59-3.95$.}
    \label{fig:streamlinesAOA4-6SpanwiseVel}
\end{figure}

As $\alpha$ increases (to $4^\circ$ and $6^\circ$; figure \ref{fig:streamlinesAOA4-6}), the separation line continues to advance, and the separation region thickens, with increasingly complex, three-dimensional surface streamline patterns beneath.  For $\alpha = 6^\circ$, the time averaged midspan flow has reattached at the higher Re $ = 6, 8 \times 10^4$, even as it leaves larger recirculation zones on either side.  The complex streamline patterns on the wall, on the other hand, have been replaced by more uniform, streamwise flow. At the lower Re $ = 2 \times 10^4$ (top row), the streamlines downstream of the separation line on the suction surface flow towards  midspan which also draws fluid from the near wall planes at $z/c=0.0075$ with flow reversal at  $x/c\approx0.8$. The flow from both wall boundaries combines with the fluid drawn from the shear layer to create a small region of apparent, quasi two-dimensional flow at midspan (as shown in figure \ref{fig:streamlinesAOA4-6SpanwiseVel}). At higher Re the wall boundary itself has returned to a uniform streamwise direction, though the circular patterns on the suction surface remain, so the influence of the wall is still felt in the interior, even up to Re $ = 8 \times 10^4$.

The streamlines in figure \ref{fig:streamlinesAOA4-6SpanwiseVel} are colored by the spanwise velocity, which reaches maximum magnitude $0.3$.  This spanwise flow decreases to net zero at the midspan symmetry plane, which is also where the LSB signature is first lost (as Re $ = 6, 8 \times 10^4$) leaving two LSB regions of shrinking size as Re increases.

\begin{figure}
    \centering
    \includegraphics[width=1\columnwidth,]{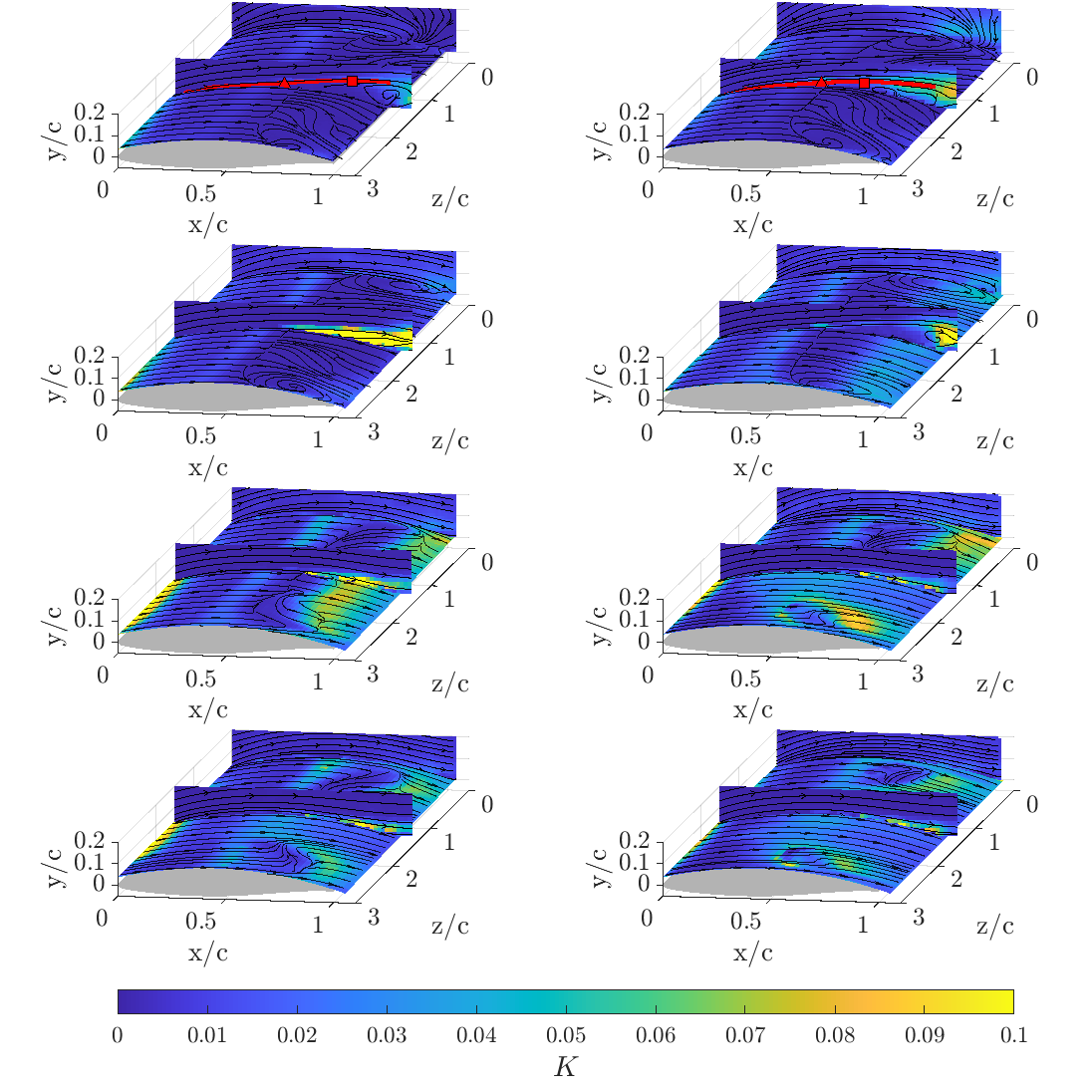}
    \caption{Kinetic energy of fluctuations, $K$, for $\alpha = 4^{\circ}$ (left) and $ 6^{\circ}$ (right). Re $= 2,4,6,8 \times 10^4$ (top to bottom). $(-)$: $\delta^*$, $(\textcolor{red}{\blacksquare})$: $\textcolor{red}{\max(H)}$.}
    \label{fig:streamlinesAOA4-6FlucEnergy}
\end{figure}

Compared with their counterparts at lower $\alpha$, the fluctuations in figure \ref{fig:streamlinesAOA4-6FlucEnergy} are higher inside the separated regions. In all cases a band of higher $K$ lies in front of the separation streamline, and a region of relatively low $K$ occurs where spanwise motions are strongest (figure \ref{fig:streamlinesAOA4-6SpanwiseVel}). The top row at Re $ = 2 \times 10^4$ shows the red line ($\delta^*$) following the shear layer with $\max(H)$ moving towards the leading edge relative to $\alpha = 0^\circ$ and $2^\circ$. The forward progression of $\max(H)$ likely indicates the onset of transition, but since the flow is still in laminar stall (an LSB has not formed) $\max(H)$ may be influenced by trailing edge vortex shedding. In other words many LSB studies \cite{kurelek:18, burgmann:08b, michelis:17} cite $\max(H)$ as the transition point, but in the absence of an LSB the $\max(H)$ relationship to transition onset may be more complicated.

\clearpage

\subsubsection{Short bubbles and flow state changes close to $\alpha_{\textrm{crit}}$}

\begin{figure}
    \centering
    \includegraphics[width=1\columnwidth,]{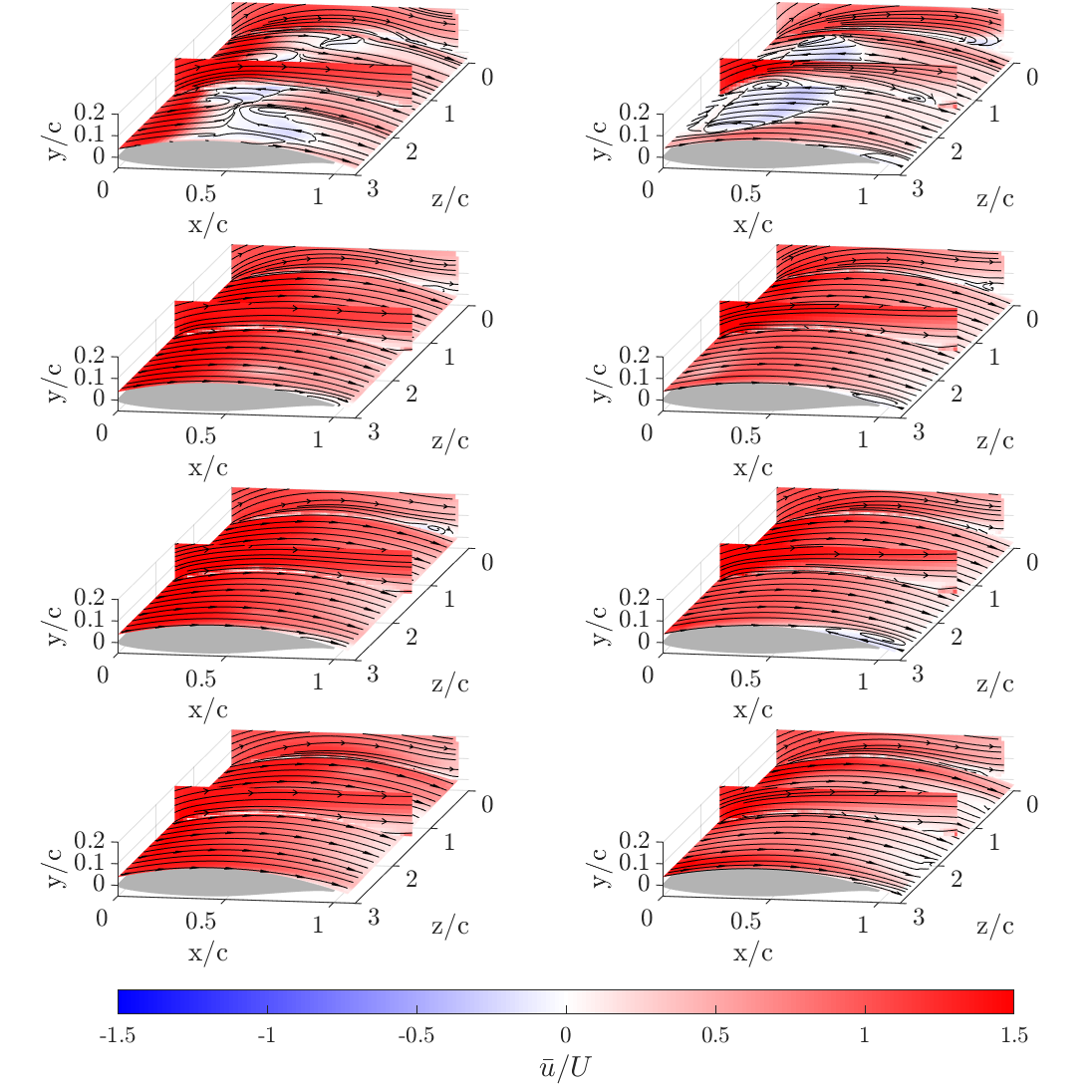}
    \caption{Time averaged streamlines and $\overline{u}$ for $\alpha = 8^{\circ}$ (left) and $ 10^{\circ}$ (right). Re $= 2,4,6,8 \times 10^4$ (top to bottom). }
    \label{fig:streamlinesAOA8-10}
\end{figure}

\begin{figure}[ht]
    \centering
    \includegraphics[width=1\columnwidth,]{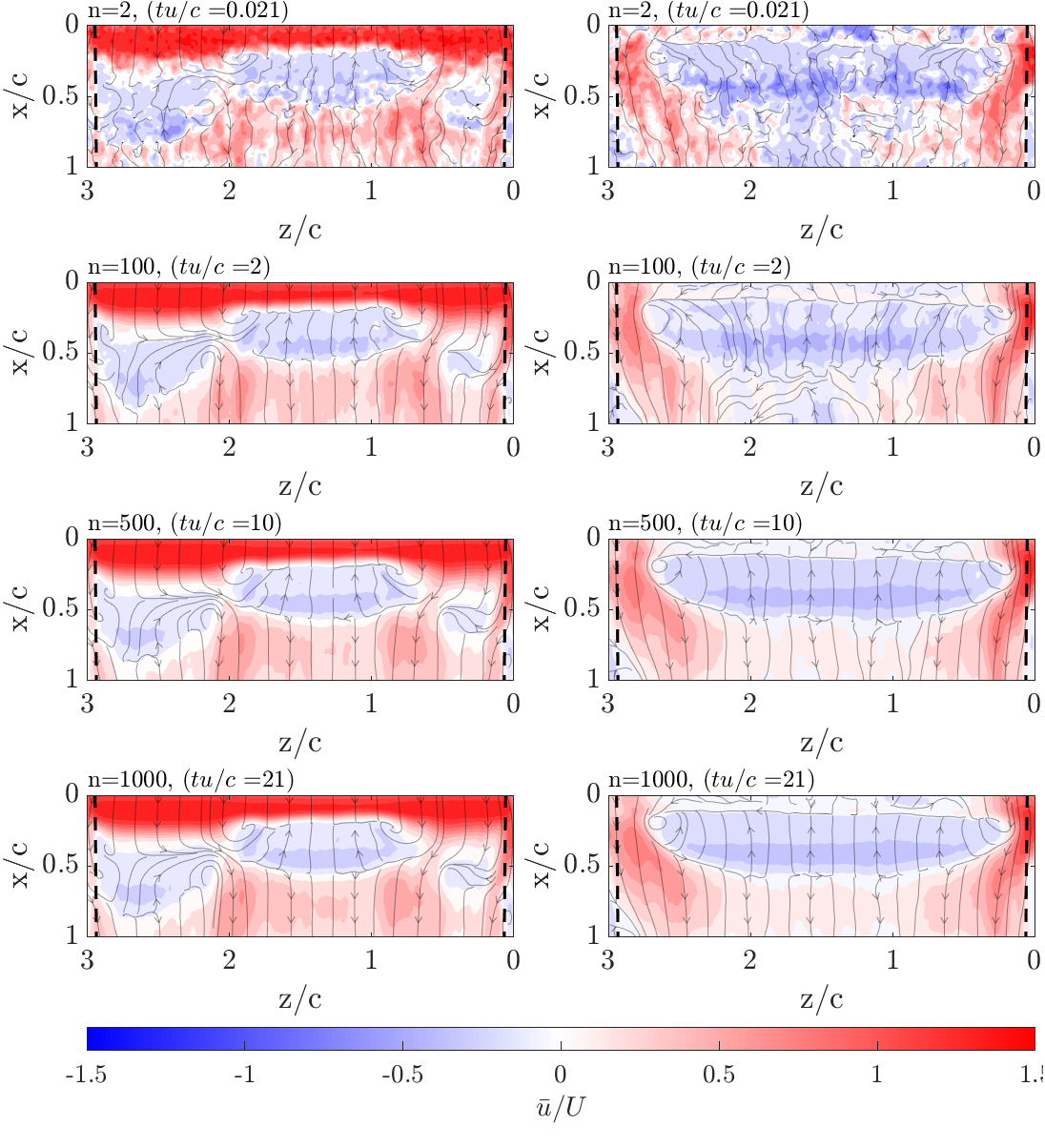}
    \caption{$\overline{u}(x,z)$ for $\alpha = 8^{\circ}$ (left) and $10^{\circ}$ (right)  for $\textrm{Re} = 2 \times 10^4$ as a function of number of samples (n) used for time averaging. ($--$): Estimate of Laminar boundary layer thickness ($\delta_{EW}(x)$) on the end walls.}
    \label{fig:streamlinesAOA8-10Re20k}
\end{figure}

At $\alpha = 8^{\circ}$ and $\alpha=10^{\circ}$ (figure \ref{fig:streamlinesAOA8-10}) the flow state has abruptly changed and flow reversal, when it occurs, is confined to the first half-chord for Re $ = 2 \times 10^4$, and then is replaced by a flow that is quite uniform and directed in the streamwise direction.  The change in flow state at some critical angle of attack $\alpha_{\textrm{crit}}$ has been noted before in experiments and simulations on the same section profile \cite{tank:21}, where the formation of a closed laminar separation bubble was associated with an increase in lift and decrease in drag. There is a recirculation region on the wall boundary near the trailing edge, and this region shrinks as Re increases from $2 \times 10^4$ to $8 \times 10^4$. At these high angles of attack, convergence of time averaged fields becomes a concern due to the shedding of large structures. Figure \ref{fig:streamlinesAOA8-10Re20k} shows the streamwise velocity contour and associated streamlines of the time averaged flow fields as a function of the number of measurement samples, $n$. The figure confirms the statistical convergence of the flow fields with consistent flow structures at $n = 500$ and $n = 1000$ for both angles of attack.

\begin{figure}
    \centering
    \includegraphics[width=1\columnwidth,]{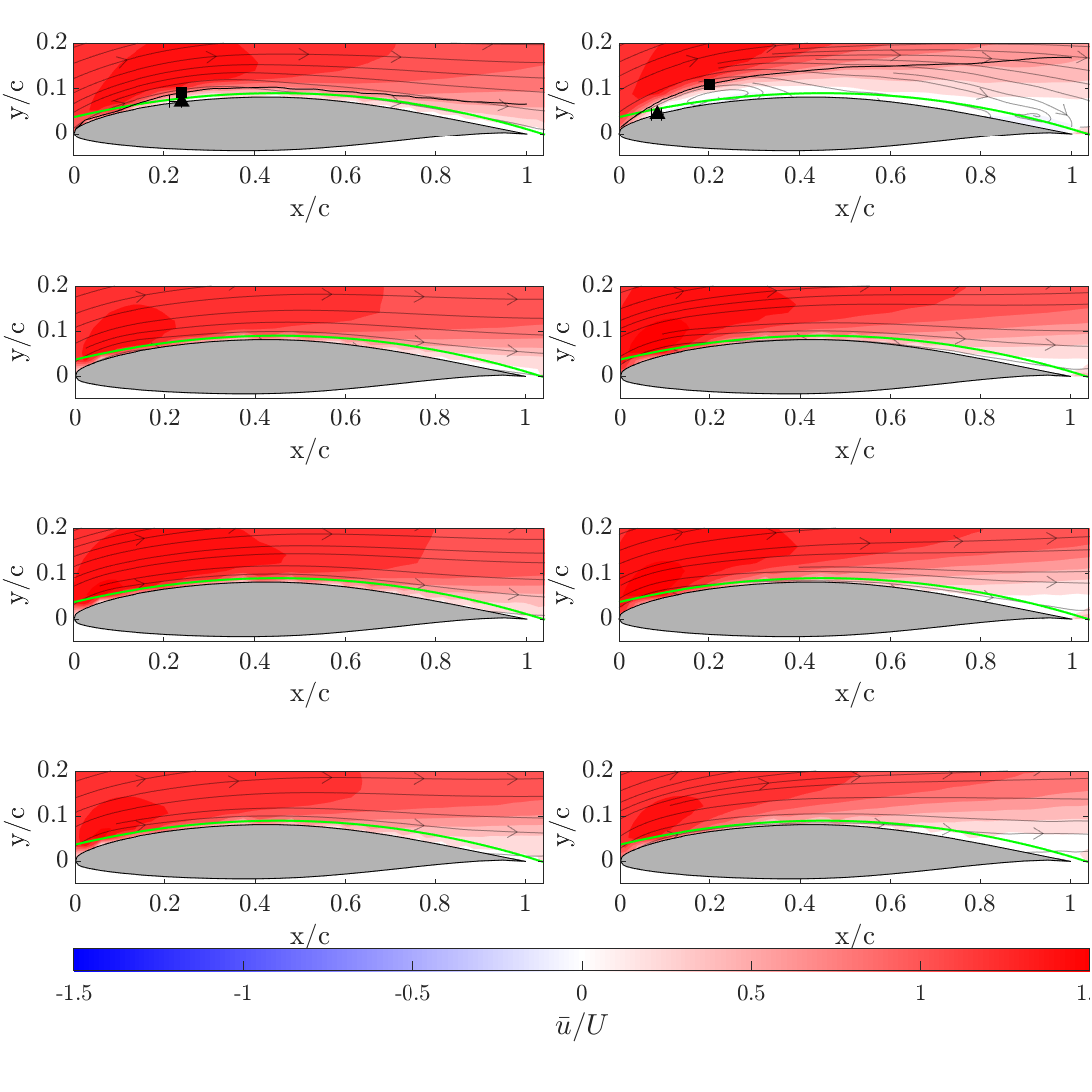}
    \caption{Time averaged streamlines and $\overline{u}(x, y)$ for $\alpha = 8^{\circ}$ (left) and $10^{\circ}$ (right) at midspan. Re $= 2,4,6,8 \times 10^4$ (top to bottom). The green line is the centerline location of the curved laser sheet over which the spanwise distributions in figure \ref{fig:streamlinesAOA0-2SpanwiseVel} are estimated. Boundary layer integral parameters $(-)$: $\delta^*$, $(\blacksquare)$: $(\max(H))$, and separation location $(\blacktriangle)$: $x_s$. $x_s$ error bars are based on $H=2.59-3.95$}
    \label{fig:midpsanAoA8-10}
\end{figure}

\begin{figure}
    \centering
    \includegraphics[width=1\columnwidth,]{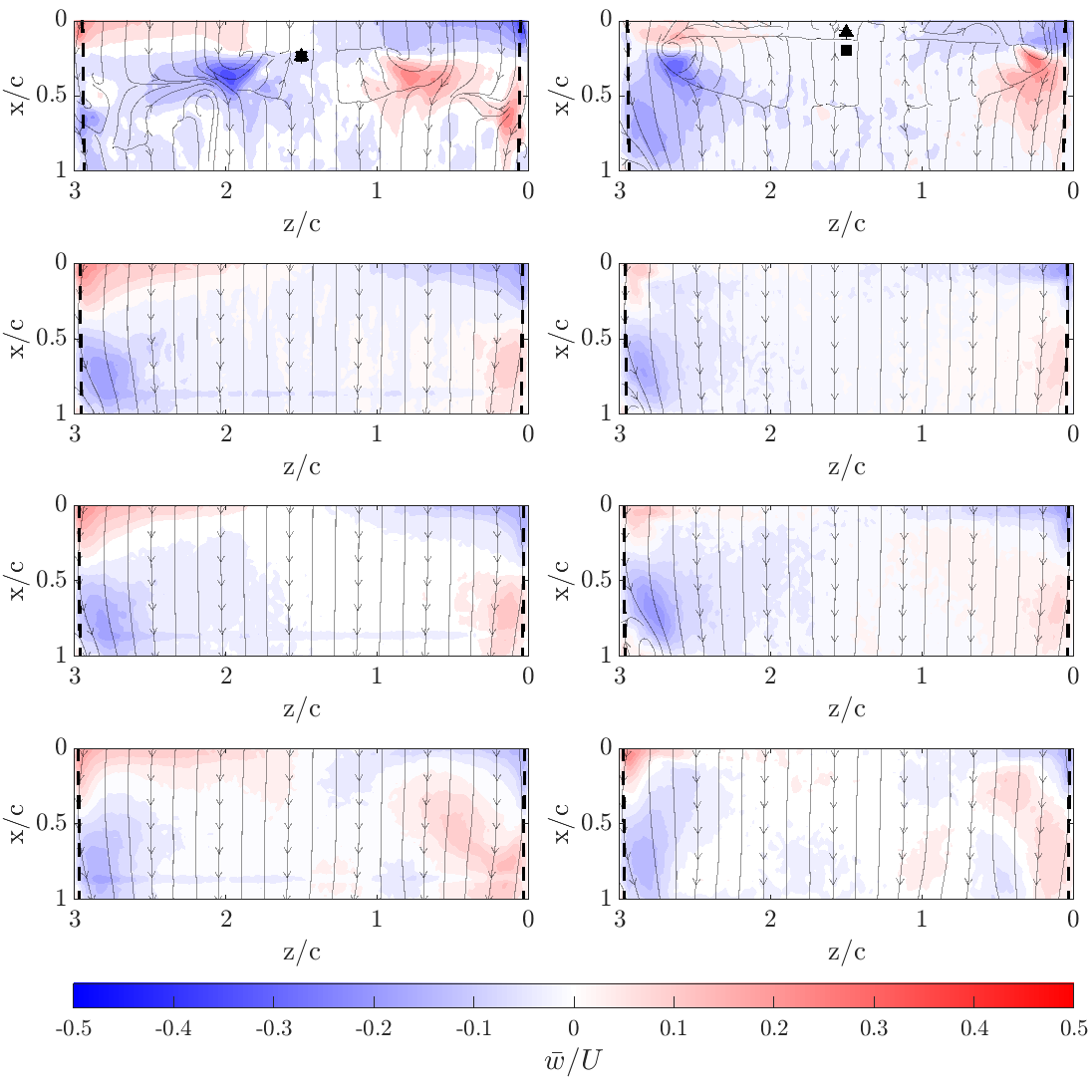}
    \caption{$\overline{w}(x,z)$ for $\alpha = 8^{\circ}$ (left) and $ 10^{\circ}$ (right). Re $= 2,4,6,8 \times 10^4$ (top to bottom). ($--$): estimate of laminar boundary layer thickness ($\delta_{EW}(x)$) on the end walls, $(\blacksquare)$: $(\max(H))$,  $(\blacktriangle)$: $x_s$. $x_s$ error bars are based on $H=2.59-3.95$.}
    \label{fig:streamlinesAOA8-10SpanwiseVel}
\end{figure}

Details of the suction surface streamlines and spanwise velocities (figures \ref{fig:streamlinesAOA8-10Re20k} \& \ref{fig:streamlinesAOA8-10SpanwiseVel}) show the Re $= 2 \times 10^4,\; \alpha = 8^\circ$, case to be almost uniquely complex.  There is a small patch of reversed flow over the central third of the wing, and at the spanwise limit near the wall junction, the lateral velocity component, $|w|$, attains its highest values. The flow is highly unsteady and nonuniform this close to $\alpha_{\textrm{crit}}$, and at $\alpha = 10^\circ$ the switch to the high lift flow state is complete with a closed recirculation region occupying most of the span. The maxima in $|w|$ are at the spanwise minimum and maximum locations of the bubble at $z/c = 0.15$ \& $2.75$. At both $\alpha = 8^\circ$ and $10^\circ$, increases in Re from its lowest value (top row) yield a flow that is, on average, uniform and attached, with the exception of small recirculating regions close to the wall junction towards the trailing edge of the blade. These recirculation regions resemble corner separation \cite{li:20, dawkins:22} and are present in $(x,y)$ slices at the end wall and $(x,z)$ slices at the suction surface for Re $\geq 2 \times 10^4$.

Figure \ref{fig:streamlinesAOA8-10Re20k} focuses on planar views of the suction surface data with streamlines and $\overline{u}(x, z)$ for Re $ = 2 \times 10^4$. At $\alpha = 8^\circ$, there are two sets of streamlines dividing the flowfield into 3 regions, each having their own recirculation zone.  At $\alpha = 10^\circ$, the recirculation zones have merged into one, which resembles an ellipse that covers most, but not all the span. The flow near the end walls is smooth and attached, other than the recirculation at the trailing-edge/wall junction.

\begin{figure}
    \centering
    \includegraphics[width=1\columnwidth,]{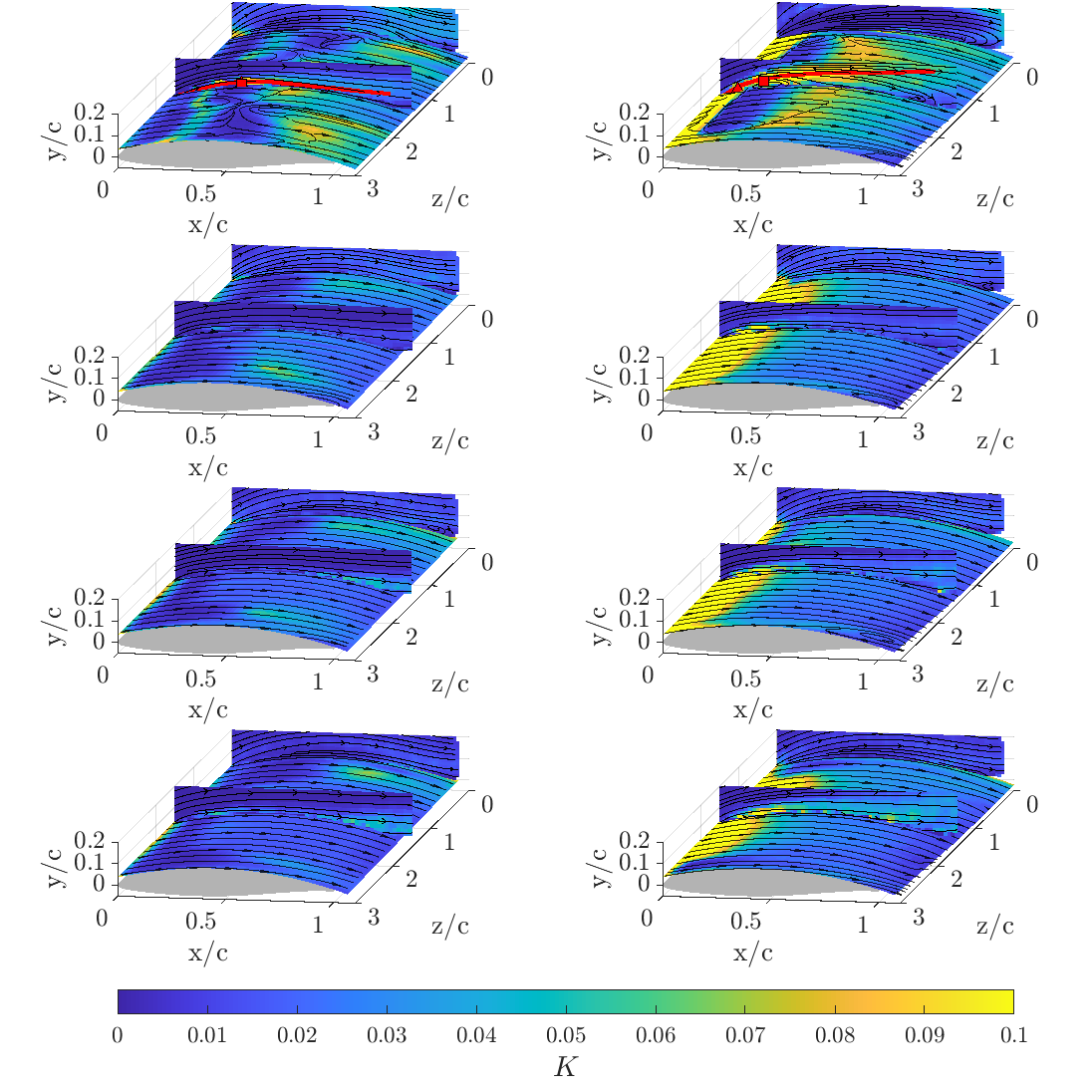}
    \caption{$K$ for $\alpha = 8^{\circ}$ (left) and $10^{\circ}$ (right). Re $= 2,4,6,8 \times 10^4$ (top to bottom). $(-)$: $\delta^*$, $(\textcolor{red}{\blacksquare})$: $\textcolor{red}{\max(H)}$.}
    \label{fig:streamlinesAOA8-10FlucEnergy}
\end{figure}

The fluctuation energies, $K$ (figure  \ref{fig:streamlinesAOA8-10SpanwiseVel}), for $\alpha=8^{\circ} $ and $10^{\circ}$ vary greatly over one chord length for the case Re $= 2 \times 10^4$ as the short laminar separation bubble moves forward, reattaching as a turbulent boundary layer, which then separates again before the trailing edge (figure  \ref{fig:midpsanAoA8-10}: top right). Beneath the bubble, the flow is laminar (indicated by dark blue), but downstream of the bubble, fluctuations peak with large yellow regions. For higher Re the time averaged streamlines still do not follow the suction surface contour all the way to the trailing edge. At $\alpha = 10^\circ$, the fluctuation magnitudes decrease with streamwise distance along the chord as the flow loses energy due to stall. The $\max(H)$ location is now firmly entrenched on top of the separation bubbles for Re $ = 2 \times 10^4$ indicating the onset of transition for both $\alpha= 8^\circ$ and $10^\circ$. $\delta^*$ at $\alpha= 8^\circ$ follows the airfoil contour down to the trailing edge, growing just slightly, while $\alpha= 10^\circ$ shows the red line diverging rapidly from the surface, downstream of the LSB, reflecting the onset of stall.

\clearpage

\section{Geometry of Separation Lines}

\begin{figure}
    \centering
    \includegraphics[width=1\columnwidth,]{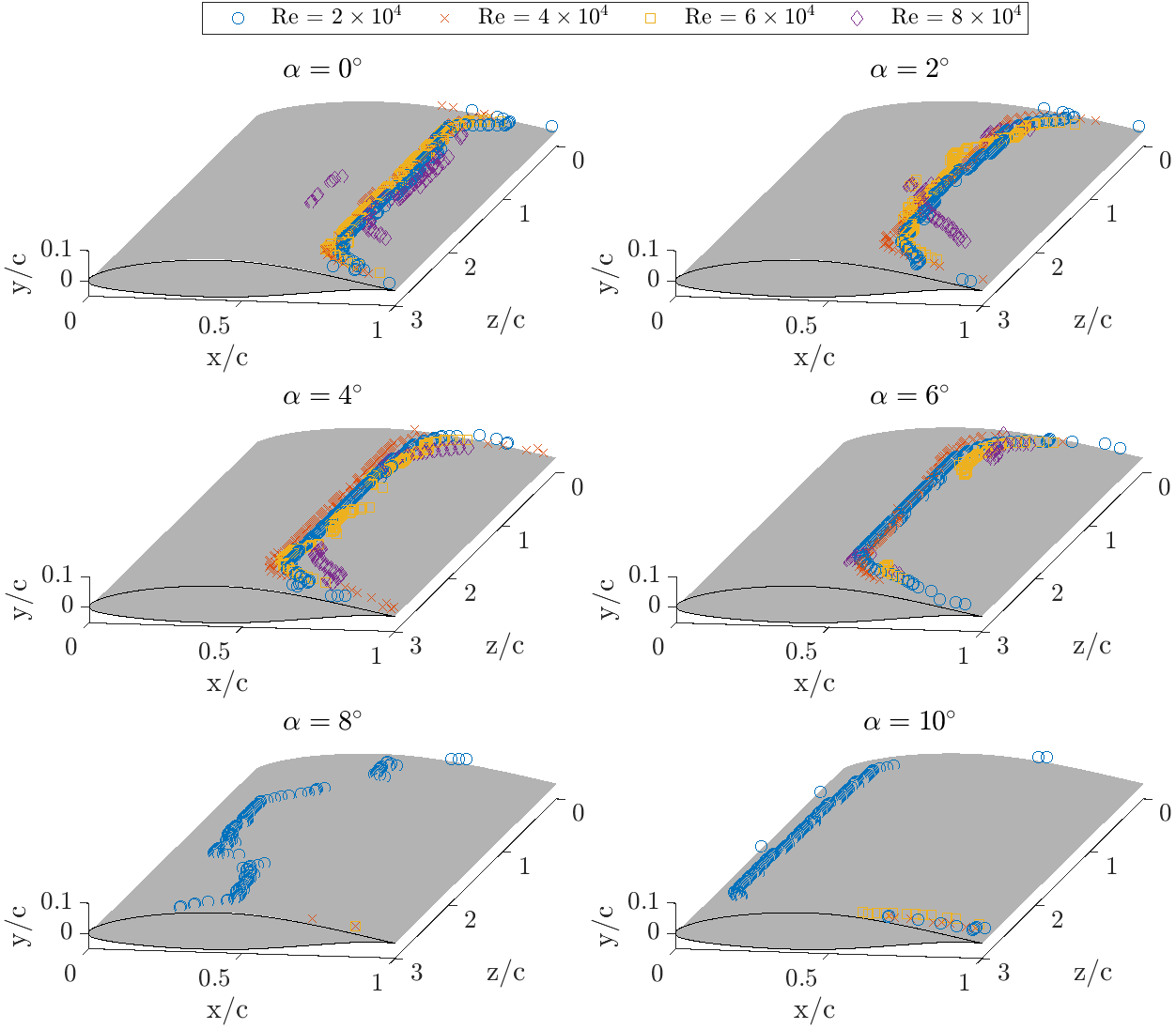}
    \caption{$\overline{u}=0$ line on the curved laser sheet.}
    \label{fig:u=0}
\end{figure}

Figure \ref{fig:u=0} shows the footprint of the $\overline{u}_0$ field at varying Re and $\alpha$.  The measured $\overline{u}_0$ location is mapped onto the curved laser sheet by interpolating between the two neighboring points in the chordwise direction where $\overline{u}$ first changes sign.  The procedure is repeated at each spanwise location.  If no reliable sign change can be found, the data are omitted.

The line of $\overline{u}_0$ is not a strong function of Re over the range of $\alpha \in [2^\circ, 4^\circ, 6^\circ, 8^\circ, 10^\circ]$.  For most of the span, up to and including $\alpha = 6^\circ$, the line is straight, curving towards the trailing edge only close to the wall boundaries.  At $\alpha = 8^\circ$ the geometry changes, the $\overline{u}_0$-line is non-uniform and has moved towards the leading edge, past the half-chord.  At the next $\alpha = 10^\circ$, it lies just behind the leading edge everywhere along the span and the transition to the short leading-edge bubble is complete. There are no data for Re $ > 2 \times 10^4$, $\alpha = 10^\circ$, except towards the trailing edge. The higher Re cases have thinner boundary layers and the separated flow may not penetrate the curved laser sheet and prevent a visualization of the $\overline{u}_0$-line, especially if the location is close to the leading edge. 

\section{Quasi-Periodic Flow Structures}

\begin{figure}
    \centering
    \includegraphics[width=1\columnwidth,]{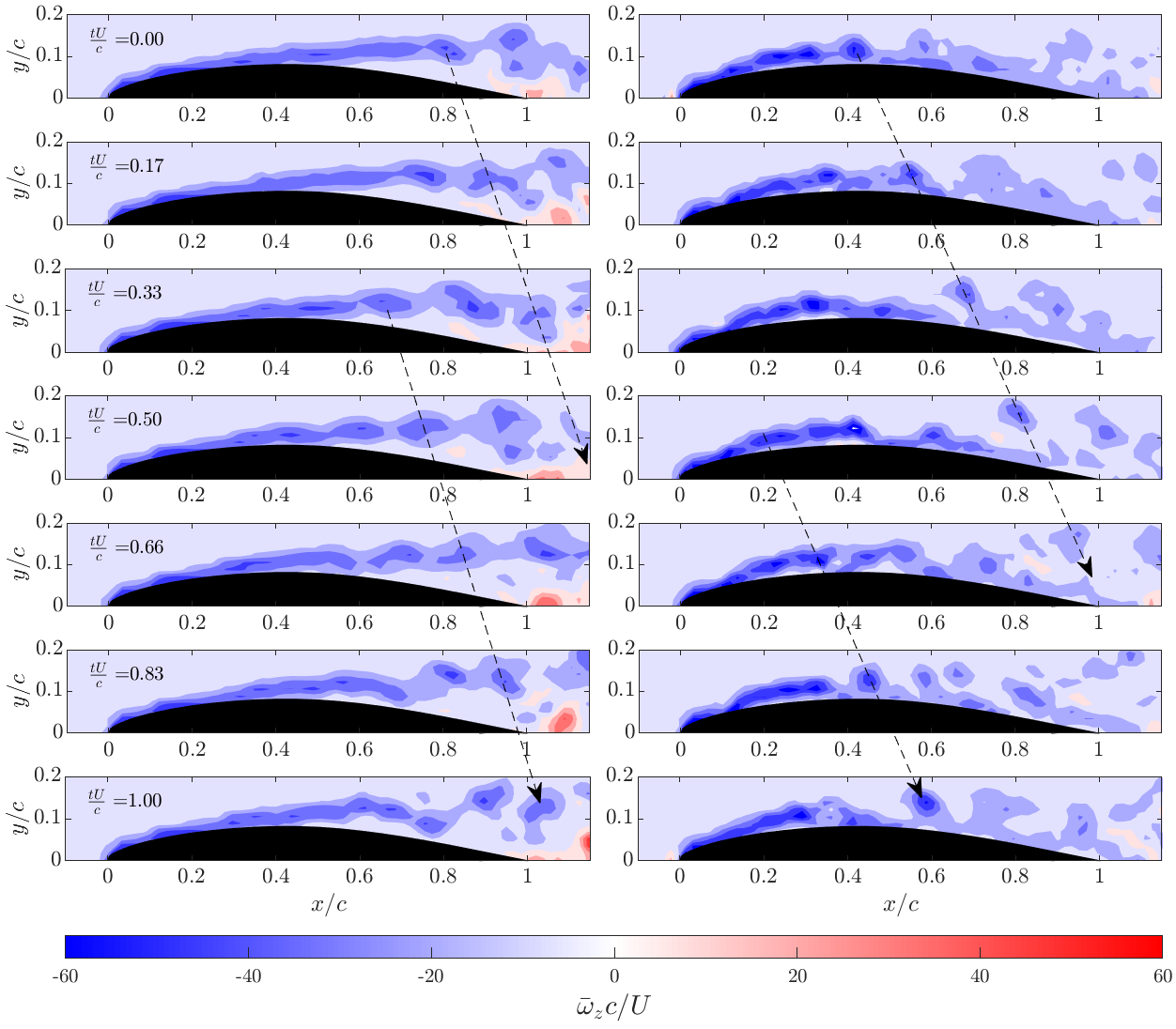}
    \caption{Instantaneous midspan snapshots of $\overline{\omega}_z$ for $\alpha=6^\circ$ (left) and $\alpha=8^\circ$ at Re $ = 2 \times 10^4$ over the course of one advection time.}
    \label{fig:Re20000Snapshots}
\end{figure}

A measure of unsteadiness in the flow with and without time averaged reattachment is shown through consecutive snapshots of the flow field in figure \ref{fig:Re20000Snapshots} over a single advection time. $\alpha =6^\circ$ illustrates the strong coherence of flow structures shed from the shear layer (which is shown as the dark blue ridge of high vorticity) and their quasi-periodicity. The shear layer at $Ut/c = 0$ appears relatively linear spanning from ($x/c$, $y/c$) = (0.4, 0.08) to (0.8, 0.1) and the shed vortices follow its trajectory. At  $Ut/c = 1$ the shear layer has curved significantly over the same coordinate span and a vortex appears close to the suction surface at ($x/c$, $y/c$) = (0.8, 0.05). This change in the shear layer shape evokes LSB flapping \cite{zaman:88} (although no definitive frequency peaks were observed for St$<1$) and bursting \cite{gaster:67}. The change in shape  is likely a result of the proximity to $\alpha_{\textrm{crit}}$ where a change in flow state occurs. 

At $\alpha =8^\circ$ the flow is reattached in all cases. The coherence and trajectory of vortices shed from the shear layer in the $\alpha =8^\circ$ case become less regular, which corresponding computations \citep{klose:25} show is accompanied by significant small-scale spanwise variations.

\begin{figure}
    \centering
    \includegraphics[width=1\columnwidth,]{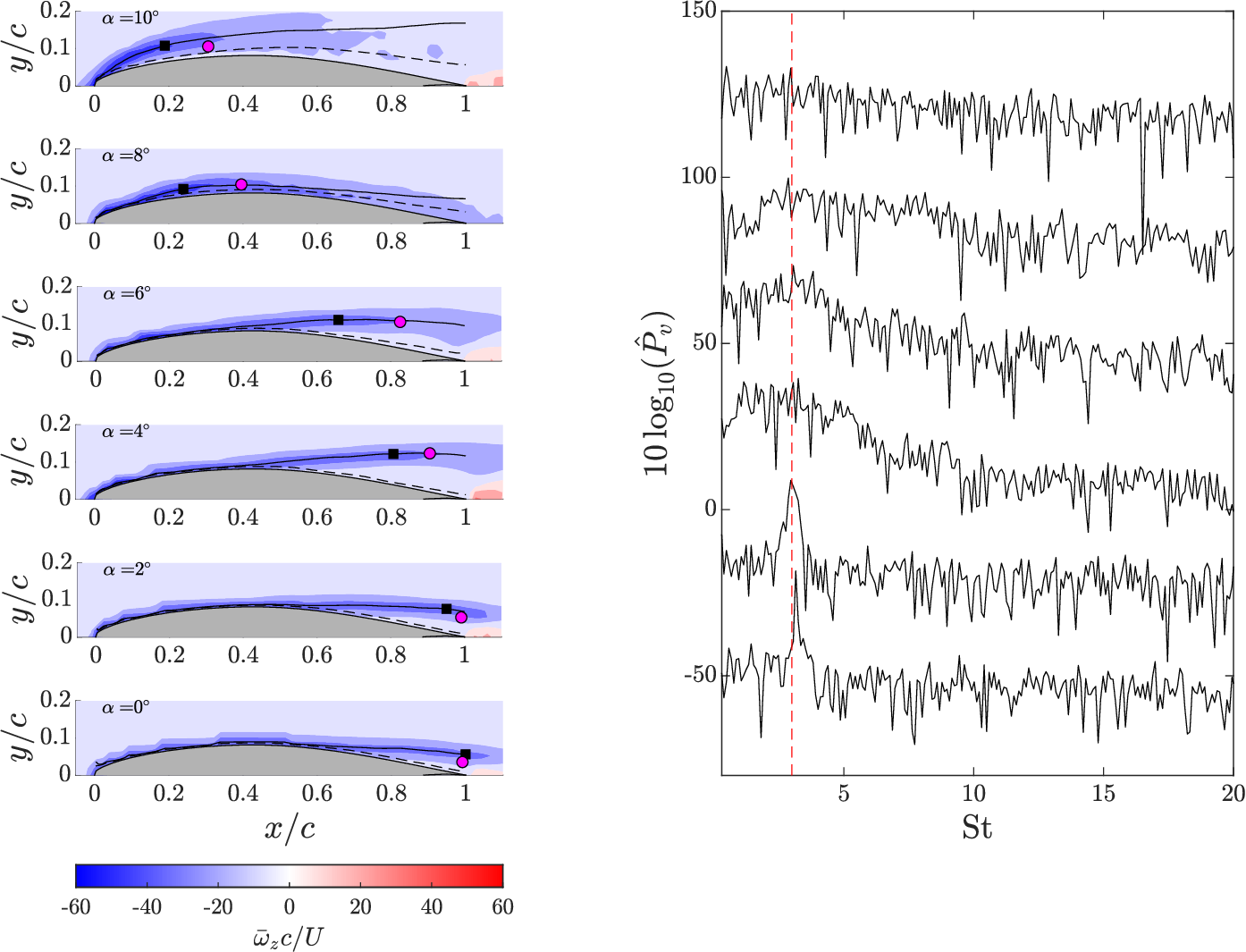}
    \caption{Midspan $\overline{\omega}_z$ at Re $ = 2 \times 10^4$ for $\alpha=0^\circ - 10^\circ$ in steps of $2^\circ$ from bottom to top, with probe location at magenta marker. $(-)$: $\delta^*$, $(--)$: $\theta$, and $(\blacksquare)$: $(\max(H))$ (left). Power spectral density of the probe $v(t)$, $(\textcolor{red}{--})$: \textrm{St=3} (right). The amplitude of each successive spectrum is stepped by 30.}
    \label{fig:Re20000PSD}
\end{figure}

The frequency content of the shear layer for fixed Re is shown in figure \ref{fig:Re20000PSD}. The location within the shear layer for $\hat{P}_v$ at each $\alpha$ is shown on the left of the figure. A persistent peak at St$ \approx 3$ occurs at each $\alpha$ and as $\alpha$ increases from $\alpha =0^\circ$ to $\alpha =10^\circ$ the relative magnitude of this peak diminishes as the two-dimensional instability in the shear layer destabilizes. The persistence of the peak at St$ \approx 3$ demonstrates the weak dependence of the shear layer shedding frequency on $\alpha$, and its reduced dominance as $\alpha$ increases reflects the decreased vortex coherence shown in figure \ref{fig:Re20000Snapshots}. The value of St $ \approx 3$ is the same as measured in experiment \cite{tank:21}, and is in reasonable agreement with computations (with periodic boundary conditions) \cite{klose:25} where St $ = 2.65$ for $\alpha = 6^\circ$.

\begin{table}
    \begin{center}
        \begin{tabular}{ c|cccccc }
         $\alpha$ & $0^\circ$ & $2^\circ$ & $4^\circ$ & $6^\circ$ & $8^\circ$ & $10^\circ$  \\
         \hline
         St$=fc/U$ & 3.15 & 2.96 & 3.24 & 3.05 & 2.86 & 2.96\\
         $\theta_s/c$ & 0.0081 & 0.0084 & 0.0080 & 0.0080 & 0.0076 & 0.0074 \\
         $(u_e/U)_s$ & 1.08 & 1.13 & 1.12 & 1.22 & 1.53 & 1.61 \\
         St$_{\theta}=f\theta_s/u_e$ & 0.024 & 0.022 & 0.023 & 0.020 & 0.014 & 0.013 \\
        \end{tabular}
    \end{center}
    \caption{Estimate of St$_{\theta} = f\theta_{s}/(u_e)_{s}$ \cite{pauley:90} from PIV at Re $ = 2 \times 10^4$}
    \label{table:St_theta}
\end{table}

The dominant frequencies from figure \ref{fig:Re20000PSD} were rescaled to St$_{\theta}=f\theta_s/u_e$ \cite{pauley:90} and tabulated in table \ref{table:St_theta} with St$_{\theta} \approx 13\times10^{-3}$ for $\alpha\geq8^\circ$ and St$_{\theta} \approx 20\times10^{-3}$ for $\alpha\leq6^\circ$. This bifurcation of St$_{\theta}$ values corresponds to the switch from a separated flow to a closed, recirculating flow with an LSB. Comparing the present measurements with values from the literature one can observe agreement at $\alpha\geq8^\circ$, but St$_{\theta}$ is larger by a factor of 1-5 than the literature for $\alpha\leq6^\circ$. These discrepancies are likely connected to the absence of an LSB for $\alpha \leq 6^\circ$, and as a result, there is a larger separation angle and $\theta_s$.  Moreover, shear-layer frequencies are significantly affected by roll-up of the pressure side boundary layer and associated shedding frequencies, so the trailing edge conditions are also influential.  The St$_{\theta}$ variations across different geometries further suggest that the pressure gradient magnitude and its streamwise distribution may affect the scaling of vortex shedding frequency through $\theta_s$ and $u_e$. Evidence of this is seen in figure \ref{fig:Re20000PSD} with very little variation of St$=fc/U$ across $\alpha$ for fixed length scales.

\begin{table}
    \begin{center}
        \begin{tabular}{ c|cccc }
         Study & St$_{\theta}\times10^{3}$  & Re$_{\theta}$ & Re$\times10^{-3}$ & $\frac{\textrm{St}}{\sqrt{\text{Re}}}\times10^{3}$ \\
         \hline
         Present (experiment) & 13-24 & 148-167 & 20 & 21\\
         Present (experiment) & - & - & 40-80 & 27\\
         \\
         Pauley et al. \cite{pauley:90} & 6.9 & 162-325 & - & - \\
         Lin and Pauley \cite{lin:96} & 5.5-8 & 79-236 & 60-200 & 10 - 17\\
         Watmuff \cite{watmuff:99} & 8.54 & 317-330 & - & -\\
         McAuliffe and Yara \cite{mcauliffe:09,mcauliffe:08} & 11 & 258-335 & - & - \\
         Kurelek \cite{kurelek:18} & 13 & 156 & 125 &  44\\
         Yarusevych \cite{yarusevych:07} & - & - & 100-150 & 31-47\\
         Michelis \cite{michelis:17} & - & - & 125 & 31\\
         Klose \cite{klose:25} & 6-13 & 70 & 20 & 15-32\\
         Burgmann and Schroder \cite{burgmann:08b} & 6-13 & 47-170 & 20-60 & 28-75\\
         Zaman and Mckinzie \cite{zaman:91} & - & - & 25-70 & 20-30 \\
        \end{tabular}
    \end{center}
    \caption{St$_{\theta}$ \cite{pauley:90} across different studies. For wind tunnel studies based on pressure-induced separation on a flat wall, the chord based Re is omitted.}
    \label{table:St_theta lit}
\end{table}

\begin{figure}
    \centering
    \includegraphics[width=1\columnwidth,]{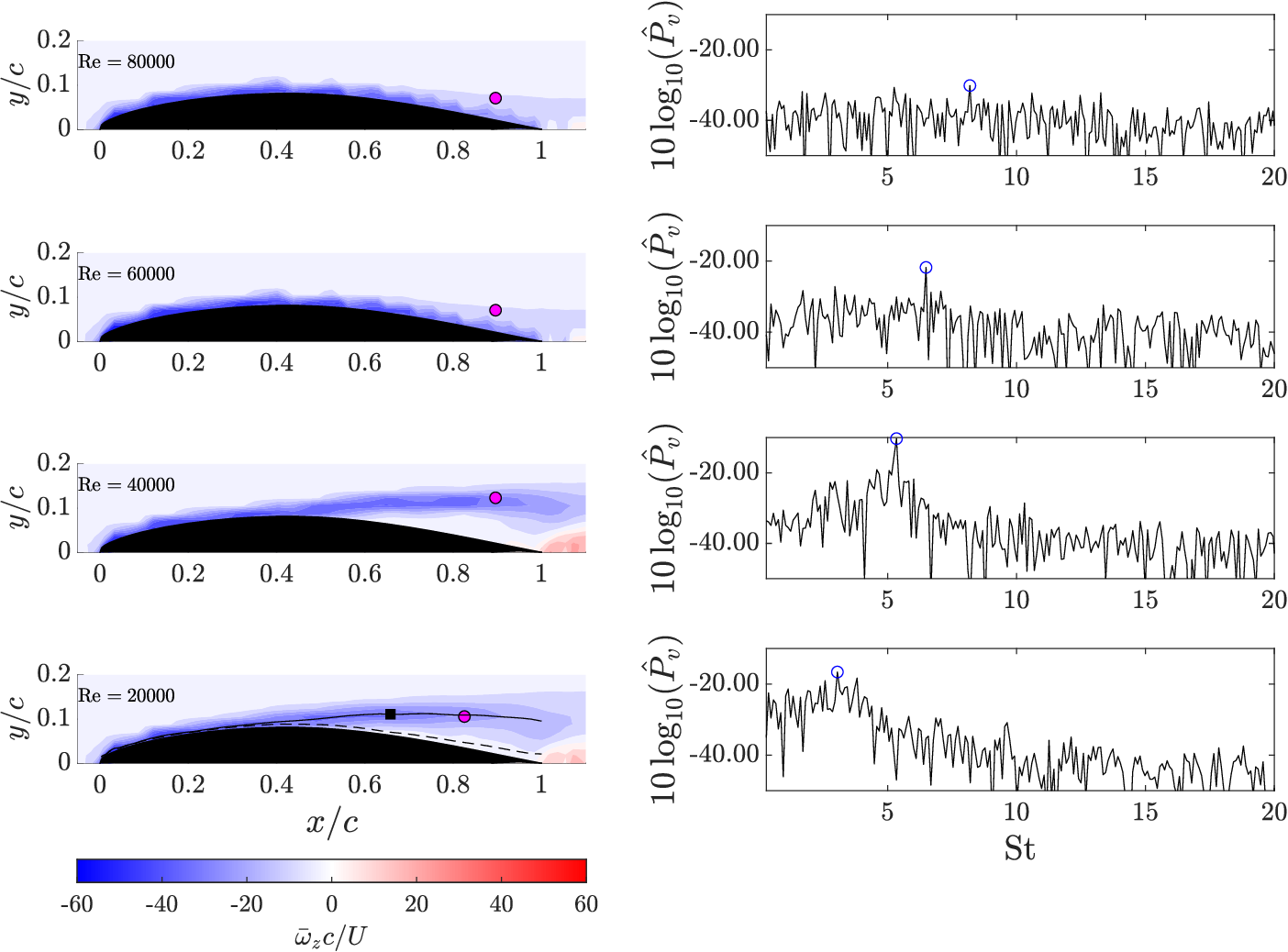}
    \caption{Midspan $\overline{\omega}_z$ at $\alpha = 6^\circ$ for Re $= 2-8 \times 10^4$ with probe location at magenta marker. $(-)$: $\delta^*$, $(--)$: $\theta$, and $(\blacksquare)$: $(\max(H))$ (left). Power spectral density of probe $v(t)$, (right). Maximum PSD frequencies identified by blue circles. Step like features in $\overline{\omega}_z$ contours at Re $= 6-8 \times 10^4$ are non-physical and a result of PIV spatial resolution}
    \label{fig:AoA06PSD}
\end{figure}

\begin{figure}
    \centering
    \includegraphics[width=0.5\columnwidth,]{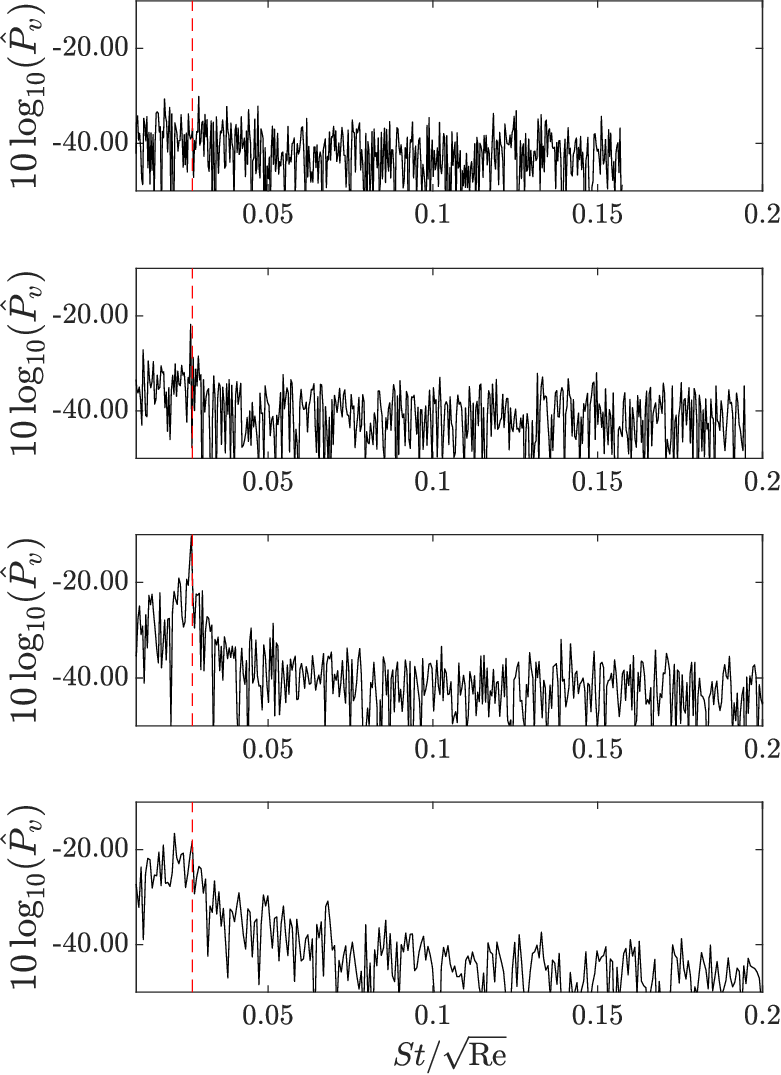}
    \caption{ Power spectral density of v(t) probe from figure \ref{fig:AoA06PSD} scaling from Zaman \& Mckinzie \cite{zaman:91}. $(\textcolor{red}{--})$: \textrm{St$\sqrt{\mathrm{Re}}=0.027$}}
    \label{fig:AoA06PSD_StRe}
\end{figure}

The variations in $St$ with Re can be seen in figure \ref{fig:AoA06PSD}. The frequency of the dominant peak increases with Re and the magnitude of the dominant peak decreases by two orders of magnitude when Re is increased from Re $= 4 \times 10^4$ to $6 \times 10^4$. This is linked to the collapse of the separated region and three-dimensional transition instabilities diffusing the energy \cite{klose:25} of the dominant two-dimensional frequency into the other frequency modes. Scaling St by $1/\sqrt{\text{Re}}$ \cite{zaman:91}, the dominant peaks collapse to a value of St/$\sqrt{\text{Re}} \approx 0.027$ shown in figure \ref{fig:AoA06PSD_StRe}. Notably, Re $= 2 \times 10^4$ does not collapse as well with the dominant frequency closer to St/$\sqrt{\text{Re}} \approx 0.021$, but a second peak at St/$\sqrt{\text{Re}} \approx 0.027$. This may be an aspect ratio effect with the end wall boundary layers stabilizing the flow and reducing the frequency at the lower Re $= 2 \times 10^4$. Otherwise, the success of a scaling based on $\sqrt{\textrm{Re}}$ suggests a viscous scaling from the initial boundary layer thickness that sets the initial conditions for shear layer dimensions. 

Drawing on this, it can be noted that if $\theta \sim 1/\sqrt{\mathrm{Re}}$ (such as for the Blasius boundary layer momentum thickness ($\theta$)) then St$_\theta$ \cite{pauley:90} is related to St$/\sqrt{\mathrm{Re}}$ \cite{zaman:91} by some factor $\sim \sqrt{n}(u_e/U)$, assuming the separation point can be approximated at some portion of the chord $x_s=c/n$:
\begin{gather*}
    \theta_{Blasiuis,s} =0.664x_s/\sqrt{\mathrm{Re_{x,s}}}\\
    x_s = c/n\\
    \textrm{St}_{\theta}=f\theta_s/u_e = \frac{0.664fx_s/u_e}{\sqrt{\mathrm{Re_{x,s}}}} = 0.664\sqrt{n}\frac{u_e}{U}\frac{\mathrm{St}}{\sqrt{\textrm{Re}}} \sim \frac{\mathrm{St}}{\sqrt{\textrm{Re}}}.
\end{gather*}

These values can be bounded as $1\leq n$ and $1\leq u_e/U\leq2$. For a trailing edge separation $n\approx1$ and $u_e/U\approx1$ resulting in St$_\theta\approx0.66(\mathrm{St}/\sqrt{\textrm{Re}})$ with $\theta_{Blasius}$. This rough estimate is in a agreement with table \ref{table:St_theta lit} and suggests that using the chord length scale may be more practical in applications as it does not require specific knowledge of separation location, boundary layer integral parameters or edge velocities. Table \ref{table:St_theta lit} illustrates that $\mathrm{St}/\sqrt{\textrm{Re}}$ values are typically between $20-50\times10^{-3}$ for airfoils with a notable exception \cite{lin:96}.

\begin{figure}
    \centering
    \includegraphics[width=0.75\textwidth,]{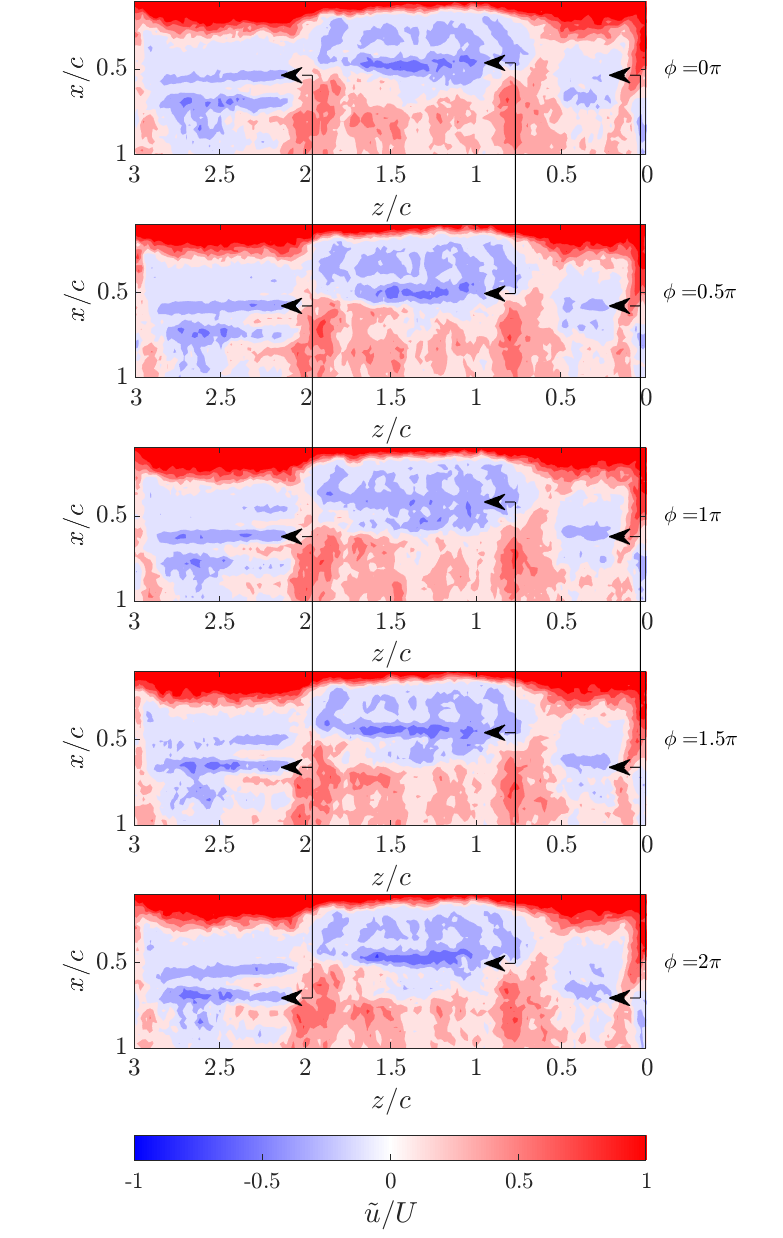}
    \caption{Contours of phase averaged flow fields $\tilde{u}$ for $\alpha=8^\circ$ at Re $ = 2 \times 10^4$ over the course of one advection time.}
    \label{fig:PhaseAverageSnapshots}
\end{figure}

Using the dominant frequency, St $ = 3$, phase-averaged snapshots of $\tilde{u}(x, z)$ are presented in figure \ref{fig:PhaseAverageSnapshots} for $\alpha=8^\circ$. The snapshots further demonstrate the unique segmentation of the flow for this particular parameter pair, with three distinct separation regions across the span. The footprints of spanwise vortex tubes which maintain their coherence from $x/c \approx 0.4$ to $x/c \approx 0.6$ convect downstream before either exiting the laser plane or breaking down to smaller structures. The onset of the recirculation region occurs downstream for the left and right regions when compared with the center region and the vortex tubes also convect further downstream. The varying recirculation lengths show that the effects of the end walls extend significantly into the flow, resulting in the formation of different separation lines with different vortex dynamics. The cellular structures are reminiscent of those observed for airfoils with wing tips \cite{Pandi:23, ribeiro:24, toppings:21} and may represent the wing root equivalent, but the circumstances under which the cells form is still not well understood and they do not appear, for example, in Toppings and Yarusevych's similar experimental \cite{toppings:22} study.

\clearpage

\section{Comparisons with Direct Numerical Simulations}

\begin{figure}
    \centering
    \includegraphics[width=1\columnwidth,]{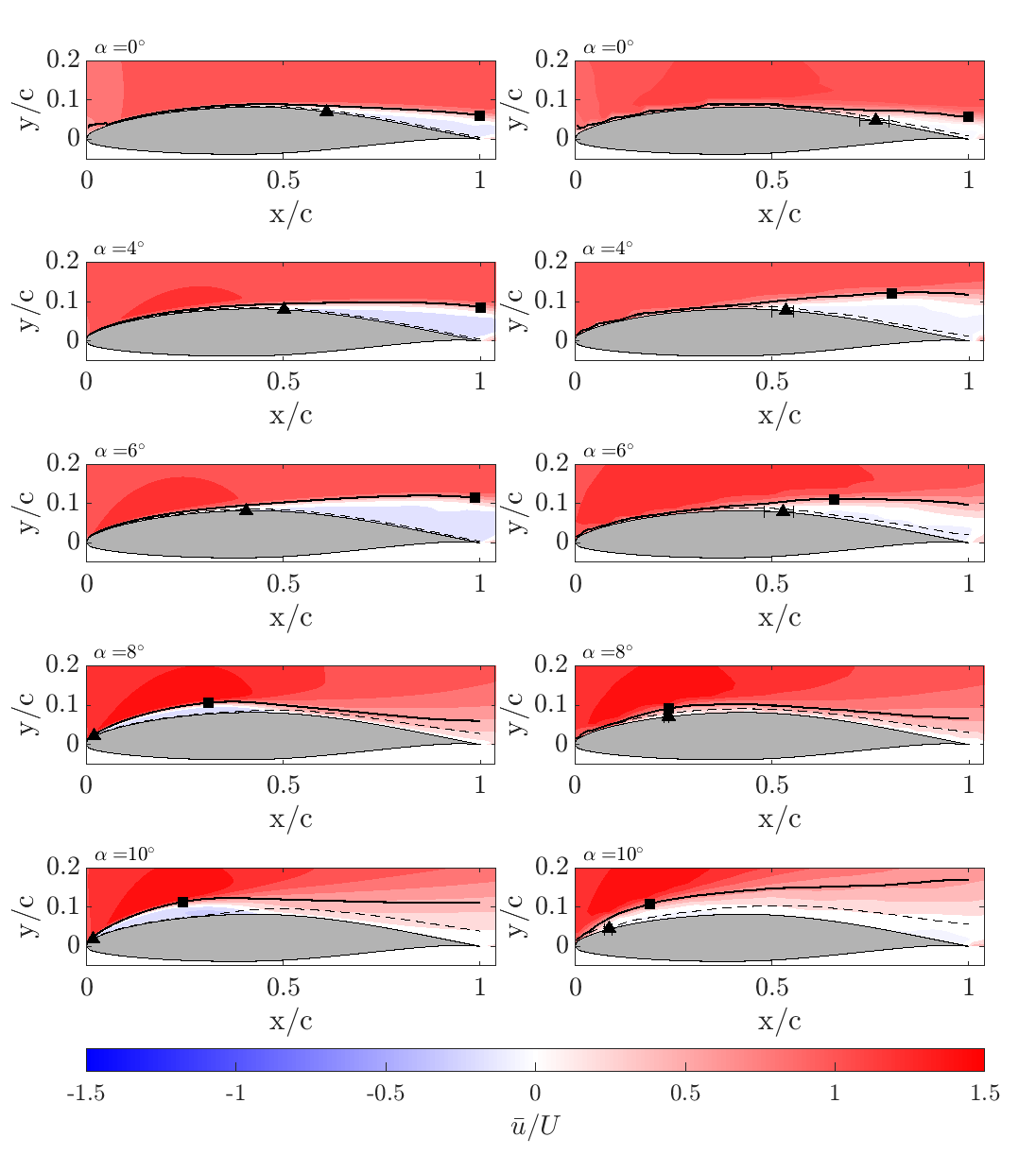}
    \caption{ $\overline{u}(x,y)$ at midspan for DNS (left) and PIV (right). Re $ = 2 \times 10^4$. $(-)$: $\delta^*$, $(--)$: $\theta$, $(\blacksquare)$: $(\max(H))$, $(\blacktriangle)$: $x_s$. $x_s$ error bars are based on $H=2.59-3.95$.
   }
    \label{fig:DNSendPeriodicComparison}
\end{figure}

\begin{figure}
    \centering
    \includegraphics[width=1\columnwidth,]{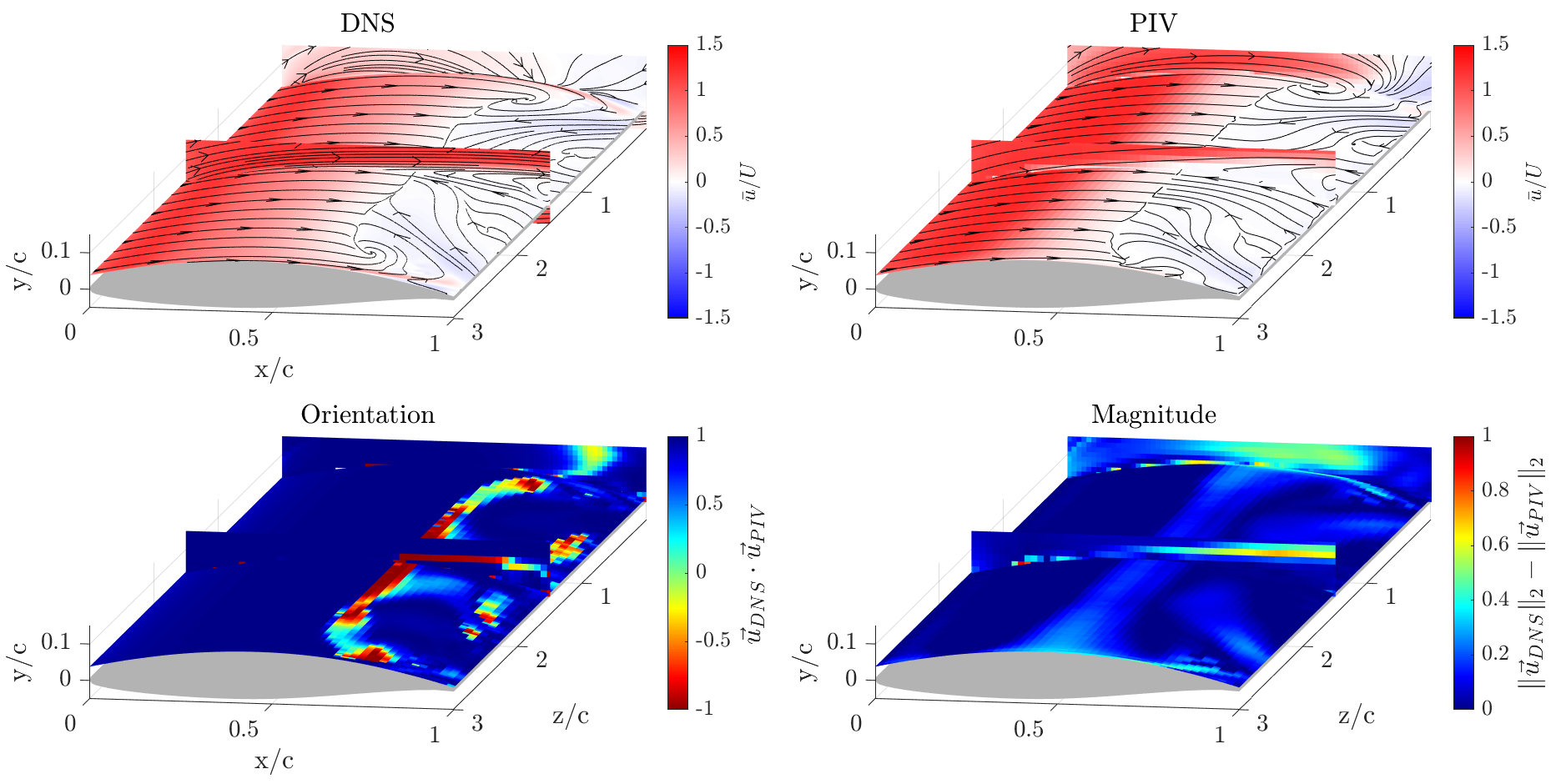}
    \caption{$\overline{u}$ on comparable slices in $(x,y)$ at midspan and at the wall, and over the suction surface in $(x,z)$ for DNS (upper left) and PIV (upper right).  The differences between the two are expressed as a normalized difference in orientation (lower left) and magnitude (lower right).
   }
    \label{fig:DNSendwallComparison}
\end{figure}

The same flow has been computed as part of the combined experimental and numerical research program at Re $=2\times10^4$. Direct numerical simulations (DNS) are currently available for AR=0.5 at $\alpha = 0^\circ, 4^\circ, 6^\circ, 8^\circ, 10^\circ$ with spanwise periodic boundary conditions and AR=3 at $\alpha = 4^\circ$ with spanwise no slip boundary conditions. For details on the computations, refer to \cite{klose:25, tank:21}. Time averaged results are compared in figure \ref{fig:DNSendPeriodicComparison} and \ref{fig:DNSendwallComparison}. The integration times of statistics for experiment and DNS were $t_{\text{exp}}\geq 10$ and $t_{\text{DNS}}\geq 8$ in advection times respectively. Separation and maximum shape factor metrics are shown in table \ref{table:sep} for experiment and DNS with spanwise periodic boundary conditions. For $\alpha\leq\alpha_{crit}$ using the criterion $H=3.95$ from the Falkner-Skan velocity profile at separation $(\beta=-0.2)$ predicts the zero skin friction point $C_f$ of the DNS data remarkably well. After the formation of the separation bubble this criterion begins to predict separation downstream of the $C_f=0$ criterion, likely related to the redistribution of the pressure gradient described by Marxen and Rist \cite{marxen:10}.

\begin{figure}
    \centering
    \includegraphics[width=1\columnwidth,]{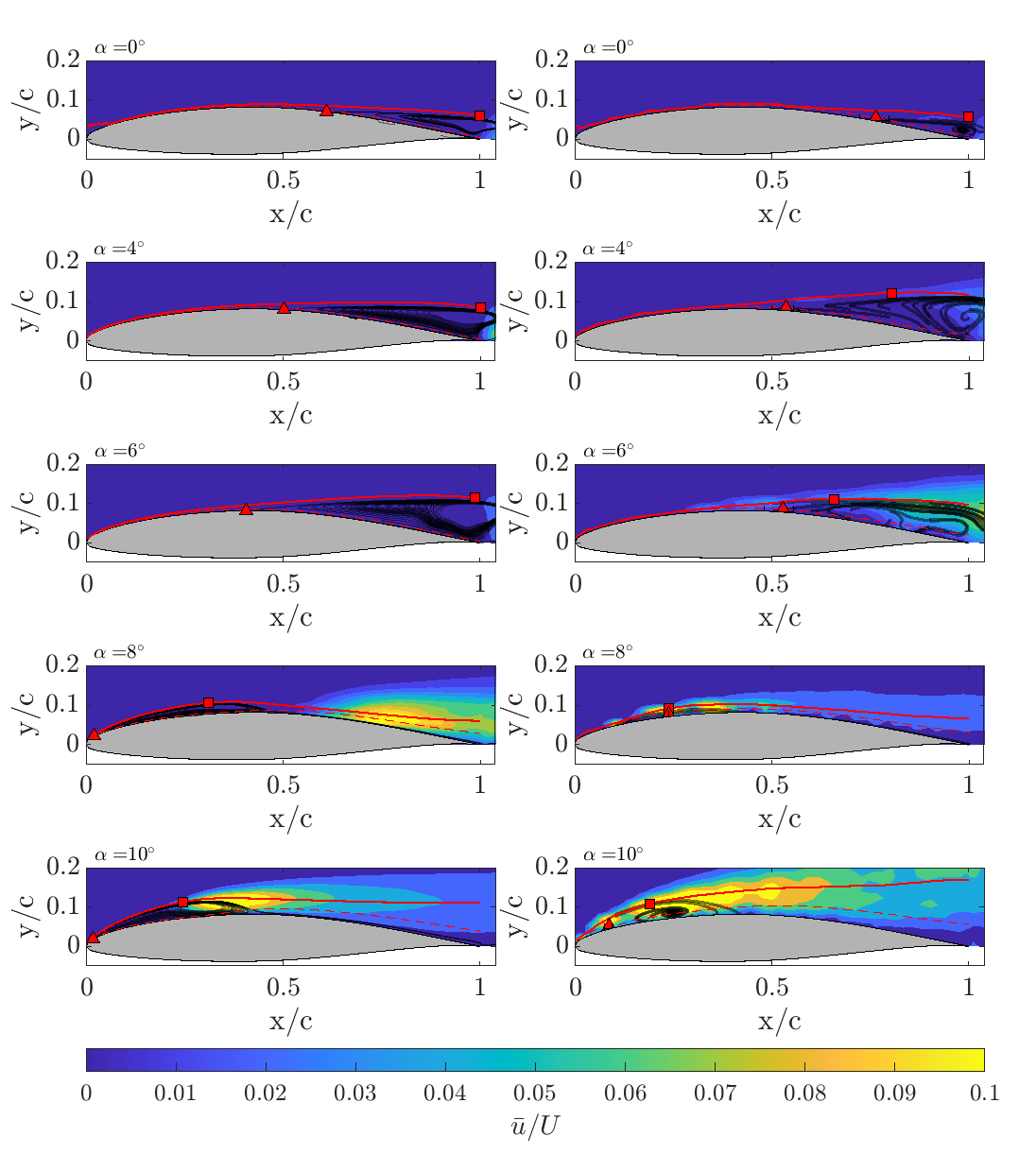}
    \caption{$K(x,y)$ at midspan for DNS (left) and PIV (right). Re $ = 2 \times 10^4$. $(\textcolor{red}{-})$: $\delta^*$, $(\textcolor{red}{--})$: $\theta$, $(\textcolor{red}{\blacksquare})$: $(\max(H))$, $(\textcolor{red}{\blacktriangle})$: $x_s$. $x_s$ error bars are based on $H=2.59-3.95$, $(-)$: streamlines within separated the region.
   }
    \label{fig:DNSTKEPeriodicComparison}
\end{figure}

\begin{table}[hb]
    \begin{center}
        \begin{tabular}{ c|ccccc }
          & $\alpha=0^\circ$ & $\alpha=4^\circ$ & $\alpha=6^\circ$ & $\alpha=8^\circ$ & $\alpha=10^\circ$  \\
         \hline
         Experiment: $x_s/c$ ($H=3.27$) & 0.76 & 0.53 & 0.53 & 0.24 & 0.086  \\
         DNS: $x_s/c$ ($H=3.95$) & 0.61 & 0.50 & 0.41 & 0.021 & 0.016  \\
         DNS: $x_s/c$ ($C_f=0$) \cite{klose:25} & 0.60 & 0.49 & 0.4  & 0.014 & 0.012
         \\
         Experiment: $\max(H)$ & 1.00 & 0.81 & 0.66 & 0.24 & 0.19  \\
         DNS: $\max(H)$ \cite{klose:25} & 1.00 & 1.00 & 0.99 & 0.32 & 0.23  \\
        \end{tabular}
    \end{center}
    \caption{Separation and maximum shape factor based on different criteria for experiment and spanwise periodic boundary condition DNS.}
    \label{table:sep}
\end{table}

\begin{figure}
    \centering
    \includegraphics[width=1\columnwidth,]{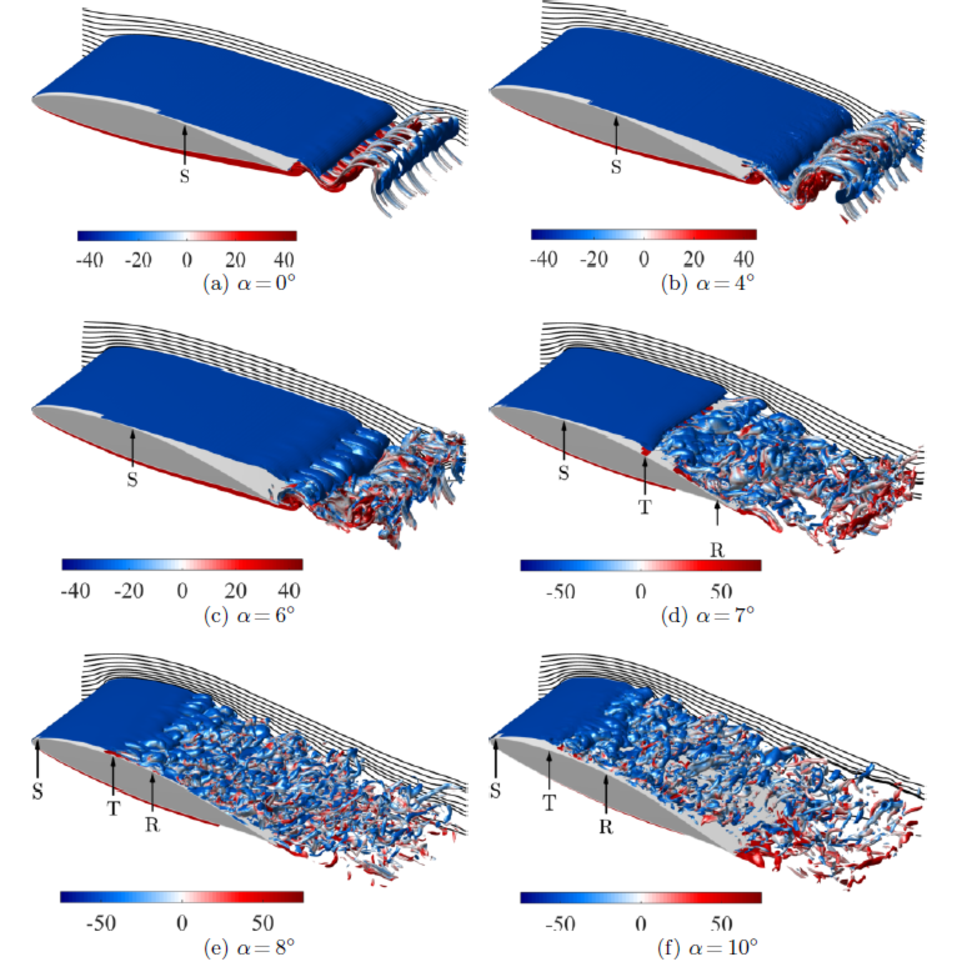}
    \caption{Iso-vorticity magnitude surfaces, colored by $\omega_z$, taken from \cite{klose:25} for $\alpha$ from $0^\circ$ (a) to $10^\circ$ (f). S, T, and R show the mean locations of separation, transition, and reattachment. 
   }
    \label{fig:Klose2024PeriodicEndWallsQCriterion}
\end{figure}

Agreement is shown in figure \ref{fig:DNSendPeriodicComparison} between the midspan slices of experiment, and simulation with periodic boundary conditions. The flow fields show similar progression of the separation region towards the leading edge with the development of LSBs for $\alpha \geq 8^\circ$. There are small differences in the height of $\delta^*$ $(-)$ and $\theta$ $(--)$. Upstream of $x_s$ this is likely from PIV resolution limitations and downstream of $x_s$ may be from the presence of higher flow perturbations in experimentation seeded through the non-zero turbulence intensities of the empty channel flow. Further evidence that ambient experimental flow conditions contribute to boundary layer growth and transition is the lagging of the DNS $\max(H)$ $(\blacksquare)$ streamwise locations behind their experimental counterparts shown in table \ref{table:sep}. Figure \ref{fig:DNSTKEPeriodicComparison} further supports this notion with magnitudes of $k$ extending further upstream in the experimental data than the DNS data. The $\alpha=6^\circ$ case in particular shows more complex streamline flow structures inside the separated region for the experimental data than the DNS data and higher $k$ values leading a $\max(H)$ location further upstream. 
Though agreement appears satisfactory in the midspan slices, Figures \ref{fig:streamlinesAOA0-2SpanwiseVel}, \ref{fig:streamlinesAOA4-6SpanwiseVel}, \ref{fig:streamlinesAOA8-10SpanwiseVel} from experiment and figure \ref{fig:Klose2024PeriodicEndWallsQCriterion} from simulation show significant non-uniformity in the spanwise direction which is not evident in figure \ref{fig:DNSendPeriodicComparison}. As an illustration of the complex three-dimensional reality that underlies selected time averaged single slices, the iso-vorticity contours in figure \ref{fig:Klose2024PeriodicEndWallsQCriterion} show a more complete description of the spatial variation downstream of separation. For $\alpha \leq 6^\circ$ the contours show spanwise uniformity up to the trailing edge with two-dimensional flapping in the near wake. For $\alpha \geq 7^\circ$ the contours show spanwise uniformity up to transition after which the flow rapidly breaks down into smaller length scale structures explaining the decreased dominance of the $\textrm{St} \approx 3$ in figure \ref{fig:Re20000PSD}.

Experiment and simulation with no slip spanwise boundary conditions are compared in figure \ref{fig:DNSendwallComparison} for $\alpha = 4^{\circ}$, Re $ = 2 \times 10^4$.  The spatial organization of the streamwise velocity fields and streamlines on the suction surface are very similar, even in the recirculation zones beneath the separation.
There are small differences in the location where streamlines on the wall-wing junction converge, and the experiment has a little higher separated region towards the trailing edge.  In both cases, the topology is the same but in slightly different positions, and as a consequence, quantitative measures of the difference in orientation and magnitude of $\Vec{\overline{u}}$ are higher at the margins of these separated regions. The inclusion of no slip boundary conditions at the spanwise limits rectifies spanwise disagreements by fixing conditions there.

\clearpage	

\section{Conclusions}

The purpose of this paper is to show how the complex, separated flows on a wall-bounded wing vary with the two controlling parameters, $\alpha$ and Re.  Laminar separation controls the wing aerodynamics at lower Re and $\alpha$. Though the separation line itself is mostly spanwise uniform, the flow fields downstream, inside the separation region are strongly three-dimensional and influenced by the wall over a significant fraction of the span. Since practical experiments must always deal with finite wings (c.f. \cite{pelletier:01}), the flow fields revealed by this study may be commonly expected in transitional-Re studies. With increasing Re, the flow at any given $\alpha$ is more likely to reattach, but the strong spanwise flows from the wall towards the midspan are always present. There is no Re where the flow is two-dimensional, or where the wall does not influence the flow at least over approximately 1/6 of the span at either end.

It has been shown before \cite{spedding:10, yang:13a, tank:21} that in the Re regime $2 \times 10^4 \leq$ Re $\leq 8 \times 10^4$, for a fixed Re there is a critical angle of attack, $\alpha_{\textrm{crit}}$ that marks an abrupt transition of a low-lift state (SI) of laminar separation without reattachment to a high-lift state (SII) of separation quickly followed by reattachment of a turbulent boundary layer. The transition is less abrupt with increasing Re, and occurs at a lower $\alpha_{\textrm{crit}}$. Resonating with two-dimensional shear layer shedding frequencies has been shown in other studies \cite{zaman:91, yang:14} to be an effective means of switching between the flow states. Here the effects of Re and $\alpha$ on shear layer shedding frequency have been explored and suitable frequencies across the parameter space have been identified for future control efforts.

The transition from SI to SII and the associated three-dimensional quasi-periodic flows are most readily observed at the lowest Re $ = 2 \times 10^4$, which acts as a test bed for the sensitivity of the separation line to small disturbances. This Re is also accessible to DNS (with no model) and the comparisons between experiment and simulation at Re $ = 2 \times 10^4$ show reasonable agreement. For future investigations into performance and flow control of wall-bounded airfoils at transitional Re, it will be important to consider the three-dimensional structure of the flow field, including the spanwise velocities.

\begin{acknowledgments}
The support from AFOSR grant FA9550-21-1-0434 under the management of Dr. Gregg Abate is gratefully acknowledged. We also thank Ari Schenkman at USC for lively discussions relating to experimental results and low Re aerodynamics.
\end{acknowledgments}
%
\bibliography{main}

\end{document}